\documentstyle[11pt]{article}

\newcommand{\bce}{\begin{center}}
\newcommand{\ece}{\end{center}}
\newcommand{\beq}{\begin{equation}}
\newcommand{\eeq}{\end{equation}}
\newcommand{\bea}{\vspace{0.25cm}\begin{eqnarray}}
\newcommand{\eea}{\end{eqnarray}}

\newcommand{\ba}{\begin{array}}
\newcommand{\ea}{\end{array}}

\newcommand{\doublespace}{
    \renewcommand{\baselinestretch}{1.6}\large\normalsize}

\def\lsim{\mathrel{\rlap{\lower4pt\hbox{\hskip1pt$\sim$}}
    \raise1pt\hbox{$<$}}}	  
\def\gsim{\mathrel{\rlap{\lower4pt\hbox{\hskip1pt$\sim$}}
    \raise1pt\hbox{$>$}}}	  

\def\Pom{{\bf I\!P}}

  \advance\oddsidemargin  by -1in
  \advance\evensidemargin by -1in
  \advance\topmargin      by -0.5in
\def\lsim{\mathrel{\rlap{\lower4pt\hbox{\hskip1pt$\sim$}}
    \raise1pt\hbox{$<$}}}         
\def\gsim{\mathrel{\rlap{\lower4pt\hbox{\hskip1pt$\sim$}}
    \raise1pt\hbox{$>$}}}         

\def\Pom{{\bf I\!P}}
\def\beq{\begin{equation}}
\def\endeq{\end{equation}}
\def\arr{\begin{eqnarray}}
\def\endarr{\end{eqnarray}}
\makeindex

 \doublespace

\begin{document}

 \large

\phantom{.}{\bf \Large \hspace{10.0cm} KFA-IKP(Th)-1993-17 \\
\phantom{.} \hspace{11cm} Landau-16/93\\
\phantom{.}\hspace{11.7cm}2 June 1993\vspace{.4cm}\\
Revised version (1 November 1993)\vspace{1.0cm}\\
}

\begin{center}
{\bf \huge The triple-pomeron regime and\\
the structure function of the pomeron  \\
in the
diffractive
deep inelastic scattering \\
at very small $x$\vspace{1.0cm}\\}
{\bf \LARGE
N.N.~Nikolaev$^{a,b}$,
and B.G.~Zakharov$^{b}$ \vspace{1.0cm} \\}
{\it
$^{a}$IKP(Theorie), KFA J{\"u}lich, 5170 J{\"u}lich, Germany
\medskip\\
$^{b}$L. D. Landau Institute for Theoretical Physics, GSP-1,
117940, \\
ul. Kosygina 2, Moscow V-334, Russia.\vspace{1.0cm}\\ }
{\LARGE
Abstract}\\
\end{center}

We discuss the diffractive deep inelastic
scattering at very small $x$ and derive the
properties of the
diffractive dissociation of virtual photons in the
triple-pomeron regime. We demonstrate that the photon-pomeron
interactions can be described by the partonic structure
function, which satisfies the QCD evolution equations, and
identify the valence and sea (anti)quark and the valence gluon
structure functions of the pomeron. The gluon structure function
of the pomeron can be described
by the constituent gluon wave function
of the pomeron.
We derive  the leading unitarization
correction to the rising structure functions at small $x$ and
conclude that the unitarized structure
function satisfies the linear evolution equations.
This resul holds even when the multipomeron exchanges
are included.
 \bigskip\\

\begin{center}
 Submitted to {\sl Zeitshrift f{\"u}r Physik C}
E-mail: kph154@zam001.zam.kfa-juelich.de
\end{center}

\pagebreak




\section{ Introduction.}


The pomeron ($\Pom$) remains one of the most misterious objects
in the high energy physics. Apart from the elastic scattering, the
exchange by pomerons describes (Fig.~1) the diffraction
dissociation of the projectile, which can be viewed as the
projectile-pomeron interaction (Fig.~1c) [1]. In the diffractive
leptoproduction at $x=Q^{2}/(Q^{2}+W^{2}) \ll 1$
one can think of the deep inelastic scattering (DIS)
on the pomeron emitted by the target nucleon [2-8].
(Here $W$ is the total energy in the photon-proton
center of mass, $W^{2}=2pq -Q^{2}$, where $p$ and $q$
are the 4-momenta of the proton and photon, and $Q^{2}=-q^{2}$
is the virtuality of the photon.)
If the diffraction dissociation is dominated by the single
pomeron exchange, which is a very strong assumption, and
if the pomeron exchange can be treated as the factorizing
particle exchnage, which also is a very strong assumption, then
one can introduce the {\sl operational} definition of the
(virtual) photon-pomeron cross section
$\sigma_{tot}(\gamma^{*}\Pom,Q^{2},M^{2})$ and of the
structure function of the pomeron
$F_{2}^{(\Pom)}(x,Q^{2})$ in terms of the
differential cross section $d\sigma_{D}/dtdM^{2}$
of the forward diffraction dissociation of virtual photons
$\gamma^{*}+p \rightarrow X+p$ ( we follow the Regge-theory
convention [1] with the substitution $M^{2} \rightarrow
M^{2}+Q^{2}$ which is natural for the DIS):
\beq
\sigma_{tot}(\gamma^{*}\Pom,M^{2})=
{16 \pi \over \sigma_{tot}(pp)}( M^{2} +Q^{2})
\left. { d\sigma_{D}(\gamma^{*}+p \rightarrow X+p)
 \over dt dM^{2}}\right|_{t=0}
\label{eq:1.1}
\endeq
and
\beq
F_{2}^{(\Pom)}(x,Q^{2})={Q^{2}\over 4\pi^{2}\alpha_{em}}
\sigma_{tot}(\gamma^{*}\Pom,M^{2})
\label{eq:1.2}
\endeq
with the corresponding Bjorken variable
\beq
x={Q^{2} \over Q^{2}+M^{2}} \, .
\label{eq:1.3}
\endeq

There has already been much work on the parton model
phenomenology of the pomeron [2-10], but the definitive
proof that the so-defined structure function of the pomeron
satisfies the conventional
Gribov-Lipatov-Dokshitzer-Altarelli-Parisi (GLDAP)
QCD evolution equations [11-13]
is as yet lacking. The definition
(\ref{eq:1.1}) for $\sigma_{tot}(\gamma^{*}\Pom)$   does
implicitly assume that the pomeron has the intercept
$\alpha_{\Pom}(0)=1$, {\sl i.e.}, the
high energy cross sections are
constant and the mass spectrum of excitation of large
masses $M$ has the $1/M^{2}$ behavior,
\beq
{1\over \sigma_{tot}(aN)}
\left.{d\sigma_{D}(a+N\rightarrow X+N)\over dt dM^{2}}\right|_{t=0}
= A_{3\Pom} {1 \over M^{2}} \, .
\label{eq:1.4}
\endeq
If the factorization relations are valid, then $A_{3\Pom}$
is expected to be a universal dimensional constant
independent of the projectile $a$. However, in QCD there
are no {\sl a priori} reasons for the factorization
relations to hold, and there are indications to the
contrary [7,8,14,15]. Furthermore, the factorization could
strongly be violated by the absorption (unitarity) effects from
the multiple pomeron exchanges in Figs.~1b,1d [6]. These multiple pomeron
exchanges cast shadow on the reinterpretation of the diffraction
dissociation in terms of the photon-pomeron interaction.
Experimentally  $\alpha_{\Pom}(0) > 1$ [16,17],
the total cross sections are rising
and the mass spectrum of the diffraction dissociation of protons
exhibits slight deviations from the $1/M^{2}$ law [18].
Furthermore, the cross section of the diffraction dissociation
of virtual photons was shown to be infrared sensitive [7-9,19].
The quantity related to the diffraction dissociation cross
section - the unitarization (shadowing, absorption)
correction to the structure
functions at small $x$ - is also infrared sensitive [19-22].
Therefore,
the possibility of introduction of the well-defined, and
well-behaved in the sense of the QCD evolution, structure
function of the pomeron and the issue of the infrared sensitivity
of the diffraction dissociation cross section
deserve further study.

In this paper we address the above problems in the framework
the light-cone $s$-channel approach to the diffractive
deep inelastic scattering at small $x=Q^{2}/(W^{2}+Q^{2})$
initiated in our previous publications [7,23,24] (for the
related early work on the $s$-channel approach to the
light-cone QED see Bjorken, Kogut and Soper  [25]).
Our strategy is to compute the high energy behavior of the
total (virtual) photoabsorption cross section and of the
diffraction dissociation cross section.
We treat the diffractive $\gamma^{*}p$ scattering in terms of
absorption of the light-cone partonic Fock components of the
virtual photon on the target proton or nucleus. One can
do so since at $x\ll 1$ the photon transforms
into these partonic Fock components at large distances
\beq
\Delta z \sim {1\over m_{N}x} \gg R_{N},R_{A}
\label{eq:1.5}
\endeq
upstream the target nucleon (nucleus). As an illustration of
the principal points of the light-cone formalism
[24,7] consider interactions of the
$q\bar{q}$ Fock state of the photon.
Because of the condition (\ref{eq:1.5})
the transverse size $\vec{\rho}$ of the $q\bar{q}$ pair and
the partition $z$ and $(1-z)$  of the (light-cone) momentum
of the photon between the quark and antiquark can be considered
frozen in the scattering process. Therefore, one can introduce
the spatial wave function of the light-cone $q\bar{q}$ Fock states
$\Psi_{\gamma^{*}}(\vec{\rho},z)$ and the dipole
cross section $\sigma(\rho)$ such that the
total photoabsorption cross section
$\sigma_{T,L}(\gamma^{*}N,x,Q^{2})$
for the (T) transverse and (L) longitudinal photons
and the forward diffraction dissociation cross section
can be calculated as the conventional quantum-mechanical
expectation value [24,7]
\beq
\sigma_{T,L}(\gamma^{*}N,x,Q^{2})=
\int_{0}^{1} dz\int d^{2}\vec{\rho}\,\,
\vert\Psi_{T,L}(z,\rho)\vert^{2}\sigma(\rho)\,\,,
\label{eq:1.6}
\endeq
\beq
\left.{d\sigma_{D}(\gamma^{*}\rightarrow X)\over dt}
\right|_{t=0}=
\int dM^{2}\,
\left.{d\sigma_{D}(\gamma^{*}\rightarrow X)\over dtdM^{2}}
\right|_{t=0}= {1 \over 16\pi}
\int_{0}^{1} dz\int d^{2}\vec{\rho}\,\,
\vert\Psi_{T,L}(z,\rho)\vert^{2}\sigma(\rho)^{2}\,\,\,.
\label{eq:1.7}
\endeq
We emphasize that the factorization of the integrands in
Eqs.~(\ref{eq:1.6},\ref{eq:1.7}) is {\sl exact},
and corresponds to the {\sl exact}
diagonalization of the scattering matrix in the
$(\vec{\rho},z)$-representation.

Interactions of the $q\bar{q}$
Fock state of the photon give the driving
terms of the structure function at small $x$ and of the
diffraction dissociation cross section. Specifically,
Eq.~(\ref{eq:1.6}) yields the
photoabsorption cross
section and the proton structure function
$F_{2}^{(N)}(x,Q^{2})$ which are
constant {\sl vs.} $x$.
Eq.~(\ref{eq:1.7}) yields the mass spectrum of
the diffractively produced states
$
d\sigma_{D}(\gamma^{*}\rightarrow X)/dM^{2}
\propto 1/(Q^{2}+M^{2})^{2}$
and can be associated with the 'valence' $q\bar{q}$
component of the pomeron [7,23,24].
Both the rise of $F_{2}^{(N)}(x,Q^{2})$ towards small $x$ and
the triple-pomeron component of the mass spectrum in the
diffraction dissociation of photons, which can be
associated with the 'sea' $q\bar{q}$ pairs and gluons
in the pomeron, are generated by interactions of the
higher, $q\bar{q}g_{1}....g_{n}$ Fock states of the photon.

The subject of this paper is a generalization of the
light-cone $s$-channel approach [24,7] to interactions
of the higher Fock states of the photon.
Our major findings
can be summarized as follows: The diffraction dissociation
cross section can indeed be factorized (Eq.~(\ref{eq:6.3}))
into the flux of
pomerons in the proton $f_{\Pom}(y)/y$, where
$y=(M^{2}+Q^{2})/(W^{2}+Q^{2})$ is a fraction of proton's
(light-cone) momentum carried by the emitted pomeron,
and the well-defined structure
function of the pomeron $F_{2}^{(\Pom)}(x/y,Q^{2})$
, which satisfies the conventional QCD
evolution with $Q^{2}$ (the definition (\ref{eq:1.1})
corresponds to the convention $f_{\Pom}(y) = 1$).
This structure function and the flux of pomerons  describe
how the naive
$\propto 1/M^{2}$ mass spectrum (\ref{eq:1.4}) is modified
by the rising hadronic
cross sections and by the QCD evolution
effects. The factorization (\ref{eq:6.3}) bears certain semblance
to the usual Regge-theory factorization despite the fact that
the Regge-theory factorization relations do not hold in deep
inelastic scattering at small $x$ and in our analysis we
never assume nor use the Regge-theory factorization. We speak
of the triple-pomeron regime just to pay a tribute to the
fact that we mostly consider $Q^{2} \ll M^{2} \ll W^{2}$,
which in the Regge-theory would have been the triple-pomeron
domain.
The infrared sensitivity of the diffraction dissociation
cross section can be reabsorbed into the initial momentum
distribution of the 'constituent' quarks, antiquarks and
gluons in the pomeron, in close similarity to the conventional
QCD evolution analysis of the proton structure function.
 We demonstrate that besides the
'valence' and 'sea' quark-antiquark components derived in
[7] (for discussion of the valence $q\bar{q}$ component of
the pomeron see also [5]),
the pomeron has the 'valence' gluon component, and present
the explicit derivation of the constituent gluon wave function
of the pomeron. The absolute normalization
of the pomeron structure function can be related to the
triple-pomeron coupling $A_{3\Pom}(Q^{2})$ Eq.~(\ref{eq:3.3.9})
, which gives the
driving term of the diffraction dissociation mass spectrum
(\ref{eq:1.4}). We shall confirm the earlier suggestion [26]
that despite being dimensionfull,
$A_{3\Pom}(Q^{2}) \propto GeV^{-2}$, this triple-pomeron coupling
only has weak dependence on
$Q^{2}$, although if $1/\sqrt{Q^{2}}$ were the
only relevant scale in the DIS, then naively one
could have expected $A_{3\Pom}(Q^{2}) \propto 1/Q^{2}$.
The fact, that one can (approximately) relate the
properties of the diffraction
dissociation in the DIS and in the
real photoabsorption, which here is proven rather than assumed,
is interesting by itself.
In [7] we gave a phenomenological estimate of the sea
structure function of the pomeron and of the diffraction
dissociation rate in terms of the triple-pomeron coupling
$A_{3\Pom}(0)$ borrowed from the real photoproduction
data [27]. The resulting predictions are in a good agreement
with the first data on the diffraction dissociation of photons
in DIS at HERA obtained recently by
the ZEUS collaboration [28].

The diffraction dissociation of virtual photons and the
unitarity (shadowing, absorption)
corrections to the rising structure functions
at small $x$ are two closely related phenomena. The
GLDAP evolution equations predict [12] very rapid rise of the
structure functions towards large $1/x$, which
conflicts the $s$-channel unitarity [19]. The $s$-channel
unitarization of the virtual photoabsorption cross section
introduces the multiple-pomeron exchanges, which could
lead to the departure of the $x$- and $Q^{2}$-dependence
of structure function from predictions of the GLDAP evolution.
We find that the unitarity corrections
are large and persist at all $Q^{2}$. The unitarity correction is
a nonlinear functional of the parton density and violates the
conventional linear relationship between the photoabsorption
cross section and the parton density.
Our
principle conclusion is that, nonetheless, the modified
evolution equations retain their linear GLDAP form as distinct from
the nonlinear equation suggested in [19] and discussed
extensively in the literature during the past decade (for a
recent review with many references see [22]).
The diagonalization of the scattering matrix in the
$(\vec{\rho},z)$-representation greatly simplifies a discussion
of the unitarization correction, as it
allows to unambiguously identify the
$s$-channel partial waves which must satisfy the unitarity bound.

This paper is organized as follows. In section 2 we
review the light-cone $s$-channel approach to deep
inelastic scattering starting with the diffractive interactions
of the two-body $q\bar{q}$ Fock state of the photon [24,7].
In section 3 we study interactions of the 3-body $q\bar{q}g$
Fock state of the photon and derive the driving term of the
triple-pomeron mass spectrum in the diffraction dissociation
of virtual photons and the driving term of the sea structure
function of the pomeron.
In section 4 we apply our formalism to derivation of the
rising structure function in the double-logarithmic limit.
The subject of
section 5 is the triple-pomeron regime of the diffraction
dissociation of photons to higher orders in the perturbative
QCD. Here we derive the structure function
and the constituent gluon wave function of the pomeron.
The latter absorbs the infared-regularization dependence
of the diffraction dissociation cross section. This completes
a derivation of the valence and sea (anti)quark and the
gluon structure functions of the pomeron, which are to be used
as an input of the GLDAP evolution of the pomeron structure
function.
The factorization properties of
the QCD pomeron and the flux of pomerons in the proton
are discussed in section 6.
In section 7 we discuss the
unitarization of the rising structure functions of the proton at
small $x$. We demonstrate that the unitarized
structure functions of the proton
still satisfy the linear GLDAP evolution
equations, as distinct from the nonlinear
Gribov-Levin-Ryskin (GLR) equations [19].
Remarkably, this conclusion holds even when the multiple pomeron
exchanges are included. In this section we also present a
brief phenomenology of the shadowing correction to the proton
structure function and comment on the treacherous path to the
interpretation of the shadowing in terms of the fusion of
partons.
In section 8 we comment on the relation between our
scenario for deep inelastic scattering at small $x$ and
the Lipatov's work on the pomeron in QCD [29]. Our main
results are summarized in section 9.

This paper is mostly devoted to the derivation of the
formalism, the numerical results for the unitarization of
the proton structure function at small $x$ [30], for
the (anti)quark and gluon distributions in the pomeron and
the structure function of the pomeron [31], for the
fusion of the nuclear partons and the nuclear shadowing
[32] will be presented elsewhere.



\section
{Deep inelastic scattering in terms of the Fock states of the
photon and the dipole cross section}



\subsection
{The $q\bar{q}$ Fock states of the photon and sea quarks in the
proton}

We are interested in DIS at $x\ll 1$
where the structure functions are dominated by the scattering
of photons on the sea quarks. The driving term of the sea
is given by the perturbative QCD diagrams shown in Fig.2.
The same diagrams can be viewed as the scattering on the
target proton of the $q\bar{q}$ Fock states of the photon.
The principle
finding of Ref.~[24] is that the corresponding contribution
to the photoabsorption cross section can be cast into the
quantum-mechanical form (\ref{eq:1.6}).
The wave functions of the $q\bar{q}$ fluctuations of the photon were
derived in [24] and read (for the related discussion in the
framework of QED
see also Bjorken et al. [25])
\beq
\vert\Psi_{T}(z,\rho)\vert^{2}={6\alpha_{em} \over (2\pi)^{2}}
\sum_{1}^{N_{f}}Z_{f}^{2}
\{[z^{2}+(1-z)^{2}]\varepsilon^{2}K_{1}(\varepsilon\rho)^{2}+
m_{f}^{2}K_{0}(\varepsilon\rho)^{2}\}\,\,,
\label{eq:2.1.1}
\endeq
\beq
\vert\Psi_{L}(z,\rho)\vert^{2}={6\alpha_{em} \over (2\pi)^{2}}
\sum_{1}^{N_{f}}4Z_{f}^{2}\,\,
Q^{2}\,z^{2}(1-z)^{2}K_{0}(\varepsilon\rho)^{2}\,\,,
\label{eq:2.1.2}
\endeq
where $K_{\nu}(x)$
are the modified Bessel functions and
\beq
\varepsilon^{2}=z(1-z)Q^{2}+m_{f}^{2}\,\,\,.
\label{eq:2.1.3}
\endeq
In Eqs.~(\ref{eq:2.1.1})-(\ref{eq:2.1.3})
$m_{f}$ is the quark mass and $z$ is the
Sudakov variable, i.e. the fraction of photon's light-cone
momentum $q_{-}$ carried by one of the quarks of the pair
($0 <z<1$). In the diagrams of Fig.~2 the color-singlet
$q\bar{q}$ interacts with the target nucleon via the Low-Nussinov
[33] two-gluon exchange, which is the driving term of the
QCD pomeron.
The interaction cross section for the color dipole of size $\rho$
is given by [24]
\beq
\sigma(\rho)={16 \over 3 }\alpha_{S}(\rho)\int
{ d^{2}\vec{k}\,V(k)[1-\exp(i\vec{k}\,\vec{\rho})]
\over
(\vec{k}^{2} + \mu_{G}^{2})^{2} }
\alpha_{S}(\vec{k}^{2}) \,\,\,.
\label{eq:2.1.4}
\endeq
In (\ref{eq:2.1.4}) $\vec{k}$ is the transverse
momentum of the exchanged
gluons, the longitudinal momentum of the exchanged gluons
is $\sim m_{_N}x$ and is
negligible at small $x$, $\,\mu_{G} \approx 1/R_{c}$
is some kind of effective mass of gluons introduced
so that color forces do not propagate beyond the confinement
radius $R_{c}\sim R_{N}$. The gluon--gluon--nucleon vertex
function $V(\vec{k})$ is related to the two-quark
form factor of the nucleon $
G_{2}(\vec{k}_{1},\vec{k}_{2})=
\langle N |
\exp\left[i(\vec{k}_{1}\cdot\vec{r}_{1}+\vec{k}_{2}\cdot\vec{r}_{2})
\right]| N\rangle$ by
\beq
V(k)=1- G_{2}(\vec{k},-\vec{k})\approx
1-{\cal F}_{ch}(3\vec{k}^{2})\,\,,
\label{eq:2.1.5}
\endeq
where ${\cal F}_{ch}(q^{2})$ is the charge form factor of the
proton. $\alpha_{S}(k^{2})$ is the running QCD coupling
which we shall use both in the momentum,
and the coordinate representations:
\beq
\alpha_{S}(k^{2})=
{4\pi \over \beta_{0}\log(k^{2}/\Lambda_{QCD}^{2})},~~~~~
\alpha_{S}(\rho)=
{4\pi \over \beta_{0}\log(1/\Lambda_{QCD}^{2}\rho^{2})}\, .
\label{eq:2.1.6}
\endeq



\subsection
{Universality of the dipole cross section and infrared
regularization}

The salient feature of the dipole cross section (\ref{eq:2.1.4}) is
its universality property [7,8]: $\sigma(\rho)$ depends only on
the size $\rho$ of the $q\bar{q}$ color dipole. The dependence
on $Q^{2}$ and the quark flavor is concentrated in the wave
functions (\ref{eq:2.1.1}) and (\ref{eq:2.1.2}).
The fundamental role of the color gauge invariance
must be emphasized. By virtue of the color gauge invariance
gluons with the wavelength $\lambda =1/k> R_{N}$ decouple
from the color-singlet nucleon. This decoupling is
taken care of by the vertex function
$V(k)$, which vanishes as $k\rightarrow 0$, and the size of
the nucleon emerges as a natural infrared regiularization:
the dipole cross section (\ref{eq:2.1.4}) is infrared-finite
even if $\mu_{G}=0$.
Similarly, the factor $[1-\exp(i\vec{k}\vec{\rho})]$ takes
care of the decoupling of gluons with $\lambda > \rho$
from the colour-singlet $q\bar{q}$ Fock state. As a result,
at small $\rho$ the cross section $\sigma(\rho)$ is perturbatively
calculable with its absolute normalization. For the nucleon target
\arr
\sigma(\rho) \approx
{4\pi\over 3 }
\rho^{2}\,\alpha_{S}(\rho)
\int_{0}^{1/\rho^{2}}
{ dk^{2} k^{2} V(k)
\over
(\vec{k}^{2} + \mu_{G}^{2})^{2} }
\alpha_{S}(\vec{k}^{2}) =
{16\pi^{2}\over 3\beta_{0}}
\rho^{2}\,\alpha_{S}(\rho)
\log\left[{1\over\alpha_{S}(\rho)}\right]=
C_{N}\rho^{2}\,\alpha_{S}(\rho) L(\rho) \, ,
\label{eq:2.2.1}
\endarr
where
\beq
L(\rho) =
\log\left[{1\over \alpha_{S}(\rho)}\right]
\label{eq:2.2.2}
\endeq
is the large parameter of the so-called {\sl Leading-Log
Approximation} (LLA) [11].

Another universal feature of $\sigma(\rho)$ is
its saturation at $\rho > R_{N},R_{c}$ because of the
confinement [24]. This is a strong-coupling regime
and
in the regime of saturation
$\sigma(\rho)$
depends on the infrared regularizations. (Following [34],
we assume the freezing strong coupling $\alpha_{S}(\rho)=
\alpha_{S}(R_{f})\sim 1$ at $\rho > R_{f}$.)
One natural
infared regularization - the size of the target proton -
enters via the vertex function $V(k)$. The other two
infrared regularizations are the effective confinement
radius $R_{c}$ which enters via the effective gluon mass
$\mu_{G}$ and the freezing point $R_{f}$ of the strong
coupling $\alpha_{S}(R_{f})\sim 1$. Evidently, by virtue
of Eq.~(\ref{eq:1.6}) the dependence on these infrared
regularizations propagates into the proton structure
function. Here we would like to emphasize that such an
infrared sensitivity of $F_{2}^{(N)}(x,Q^{2})$ is old news:
in the conventional QCD--improved parton model
this dependence is hidden in the
parameterization of the input parton distributions at the low
factorization scale $Q_{0}^{2}$. In our light-cone $s$-channel
approach we rather calculate the structure function in
terms of the dipole cross section, reducing drastically
the number of the infrared parameters. Furthermore,
the crude test of the large-$\rho$ behavior of the dipole
cross section $\sigma(\rho)$ is provided by the hadronic
cross sections. For instance, the pion-nucleon total cross
section can be evaluated as
\beq
\sigma_{tot}(\pi N)=
\int_{0}^{1} dz\int d^{2}\vec{\rho}\,\,
\vert\Psi_{\pi}(z,\rho)\vert^{2}\sigma(\rho)\,\,,
\label{eq:2.2.3}
\endeq
which roughly reproduces the observed value of
$\sigma_{tot}(\pi N)$ at moderate energies [7,8,14,15,30].
Once the constraint (\ref{eq:2.2.3}) has been enforced, the
predictions for DIS become
to a large extent parameter-free.
Such a minimal-regularization approach leads to a viable
description of the absolute value of $F_{2}^{(N)}(x,Q^{2})$
[23,24,30,35], of the longitudinal structure function
$F_{L}(x,Q^{2})$ [36] and of the nuclear shadowing in DIS
[23,24,35], of the gluon distribution in the proton [37],
of the excitation of charm in the muon and neutrino
scattering [38] and roughly reproduces the total cross
section of the real photoabsorption [7,8].
This minimal-regularization approach is not imperative,
though, and all the results to be presented below could
easily be reformulated in terms of more familiar
parameterizations of
the input parton distributions in the proton and pomeron
at the expense of losing the predictive power.



\subsection
{Connection with the QCD evolution equations}

At small $Q^{2}$ the transverse size of the $q\bar{q}$
Fock states of the photon shall be given by the Compton
wavelength of the quark $1/m_{f}$. For the heavy flavors
(charm,...) this is the perturbatively small size. For
the light flavors it is natural to ask that also the quarks
do not propagate beyond the confinement radius. With
the natural choice $m_{u,d} \sim \mu_{\pi} \sim 1/R_{c}$
Eq.~(\ref{eq:1.6}) with the wave function (\ref{eq:2.1.1})
reproduces the real photoabsorption cross section [7,8]. At
larger $Q^{2} \gg m_{f}^{2}, \mu_{G}^{2}, R_{N}^{-2}$
the conventional QCD-improved
parton model description is recovered.
Indeed,
let us calculate the $Q^{2}$ dependence of the cross
section (\ref{eq:1.6}). At large $Q^{2}$ the leading
contribution to $\sigma_{T}(\gamma^{*}N)$ comes from the
$K_{1}(\varepsilon\rho)^{2}$ term in Eq.~(\ref{eq:2.1.1}).
Making use of the properties of the modified  Bessel functions,
after
the $z$-integration one can write
\arr
\sigma_{T}(\gamma^{*}N,x,Q^{2})=
\int_{0}^{1} dz\int d^{2}\vec{\rho}\,\,
\vert\Psi_{T}(z,\rho)\vert^{2}\sigma(\rho)\propto
{1\over Q^{2}} \int_{1/Q^{2}}^{1/m_{f}^{2}}
{d \rho^{2} \over \rho^{4} }\sigma(\rho) \nonumber\\
\propto {1\over Q^{2}} \int_{1/Q^{2}}^{1/m_{f}^{2}}
{d \rho^{2} \over \rho^{2} }
\alpha_{S}(\rho)
\log\left[{1\over \alpha_{S}(\rho)}\right]\propto
{1\over Q^{2}}{1\over 2!}
L(Q^{2})^{2}  \,.
\label{eq:2.3.1}
\endarr
We find the scaling cross section $\propto 1/Q^{2}$ times
the LLA scaling violation factor, with one power of
$L(Q^{2})=\log\left[1/\alpha_{S}(Q^{2})\right]$ per QCD loop,
which is a starting point of the derivation [11-13] of the
QCD evolution equations.
Notice, that the factor $1/Q^{2}$ in Eq.~(\ref{eq:2.3.1}),
which provides the Bjorken scaling,  comes
from the probability of having the $q\bar{q}$ fluctuation
of the highly virtual photon.
There is a finite, and also scaling, contribution to
$\sigma_{T}$ from the region of $\rho^{2} < 1/Q^{2}$:
\arr
\Delta\sigma_{T}(\rho^{2}<1/Q^{2},x,Q^{2})
\propto
\int_{0}^{1/Q^{2}}
{d \rho^{2} \over \rho^{2} }\sigma(\rho)
\propto \int_{0}^{1/Q^{2}}
d \rho^{2}
\,\alpha_{S}(\rho)
L(\rho)\propto
{1\over Q^{2}}\alpha_{S}(Q^{2})
L(Q^{2})  \,.
\label{eq:2.3.2}
\endarr
This is the
$\sim
\alpha_{S}(Q^{2})/
L(Q^{2})$ correction to the
LLA cross section (\ref{eq:2.3.1}). Notice the strong
ordering in the LLA
cross section:
\beq
1/Q^{2} < \rho^{2} < 1/m_{f}^{2} \, .
\label{eq:2.3.3}
\endeq
The QCD scaling violations are (logarithmically) dominated
by $\rho^{2} \sim 1/Q^{2}$.
Similar analysis gives the longitudinal cross section
\arr
\sigma_{L}(\gamma^{*}N,x,Q^{2})=
\int_{0}^{1} dz\int d^{2}\vec{\rho}\,\,
\vert\Psi_{L}(z,\rho)\vert^{2}\sigma(\rho)\propto
{1\over Q^{4}} \int_{1/Q^{2}}^{1/m_{f}^{2}}
{d \rho^{2} \over \rho^{6} }\sigma(\rho) \nonumber\\
\propto {1\over Q^{4}} \int_{1/Q^{2}}^{1/m_{f}^{2}}
{d \rho^{2} \over \rho^{4} }
\alpha_{S}(\rho)
\log\left[1/\alpha_{S}(\rho)\right]\propto
{1\over Q^{2}}\alpha_{S}(Q^{2})
L(Q^{2}) \,,
\label{eq:2.3.4}
\endarr
which is completely
dominated by $\rho^{2} \sim 1/Q^{2}$ (for the more
discussions on this point see [36]).



\subsection
{Diffraction excitation of the $q\bar{q}$ Fock state of the
photon and the 'valence' $q\bar{q}$ component of the
pomeron}

The shape of the mass spectrum from the diffraction excitation
of the $q\bar{q}$ Fock state of the
photon (Fig.3) can be estimated undoing the $z$-integration in
Eq.~(\ref{eq:1.7}).  Firstly, we  note  that the
diffraction dissociation cross section
is dominated by large $\rho^{2} \sim R_{c}^{2}, m_{f}^{-2}$ [7]:
\arr
\left.{d\sigma_{D,T}(\gamma^{*}\rightarrow X)
\over dt}\right|_{t=0}= {1\over 16\pi}
\int_{0}^{1} dz\int d^{2}\vec{\rho}\,\,
\vert\Psi_{T}(z,\rho)\vert^{2}\sigma(\rho)^{2}\nonumber\\
\propto
{1\over Q^{2}} \int
d \rho^{2} \left[\sigma(\rho) \over \rho^{2}\right]^{2}
\exp(-2m_{f}\rho)
\propto
{1\over Q^{2}} \int^{1/m_{f}^{2}}
d \rho^{2} \propto
{1\over Q^{2}m_{f}^{2}}
\label{eq:2.4.1}
\endarr
Secondly, the invariant mass
squared of the $q\bar{q}$ system equals
\beq
M^{2}={\vec{k}_{q}^{2} + m_{f}^{2} \over z(1-z)} \, ,
\label{eq:2.4.2}
\endeq
where $\vec{k}_{q}$ is the transverse momentum of the (anti)quark
of the pair. For the crude estimation of the mass spectrum
at $M^{2} > Q^{2}$
one can undo the $z$ integration in (\ref{eq:1.7},\ref{eq:2.4.1})
making
use of $k^{2} \sim 1/\rho^{2}\sim m_{f}^{2}$,
so that $M^{2} \sim 1/z\rho^{2}$
and
\beq
dz \sim {1\over \rho^{2}} {dM^{2}\over M^{4}}\,\, ,
\label{eq:2.4.3}
\endeq
which gives (for the more detailed
derivation see [7])
\arr
\left.{d\sigma_{D,T}(\gamma^{*}\rightarrow X)
\over dM^{2}dt}\right|_{t=0}\propto
{1\over (Q^{2}+M^{2})^{2}}\int^{1/m_{f}^{2}}
d \rho^{2} \propto
{1\over m_{f}^{2}
 (Q^{2}+M^{2})^{2}}
\label{eq:2.4.4}
\endarr
Notice the strong flavor dependence of the diffraction
dissociation cross section, Eqs.~(\ref{eq:2.4.1}),
(\ref{eq:2.4.4}). For the excitation of heavy flavors the
diffraction dissociation cross section is perturbatively
calculable, for the excitation of light flavors it is
evidently infrared-regularization dependent.
(We shall encounter such an infrared-regularization
sensitivity of the diffraction dissociation cross section
over and over again. We shall demonstarte that, however,
this infrared sensitivity can be reabsorbed into the
normalization of the input structure function of the pomeron,
which by itself will be proven to satisfy the GLDAP evolution.)
The corresponding contribution to the structure function
of the pomeron has the form [7]
reminiscent of the valence structure function of the proton:
\beq
F_{2}^{(\Pom)}(x,Q^{2}) =
{4 Q^{2}\over \pi\alpha_{em}\sigma_{tot}(pN)}
(M^{2}+Q^{2})\left.
{ d\sigma_{D}(\gamma^{*} \rightarrow q+\bar{q}) \over
dt\, dM^{2}}\right|_{t=0} \propto x(1-x)
\label{eq:2.4.5}
\endeq
(The convention (\ref{eq:1.1}) for the photon-pomeron cross
section differs from that used in Ref.~[7] by the factor
$(M^{2}+Q^{2})/M^{2}$.)
Thefore, the diffraction excitation of the $q\bar{q}$ Fock
state of the photon can be associated with DIS on the 'valence'
$q\bar{q}$ component of the pomeron (see also Ref.~[5]).



\subsection
{Rising structure functions and higher Fock states of the photon}

The diagrams of Fig.~2 can also be reinterpreted as the
Bethe-Heitler production of the $q\bar{q}$ pair by the
photon-gluon fusion $\gamma^{*}g \rightarrow q\bar{q}$.
The conventional Weizs\"acker-Williams formula for this
Bethe-Heitler cross section reads
\beq
\Delta\sigma_{tot}(\gamma^{*}N,W^{2},Q^{2}) = \sum_{1}^{N_{f}}
\int_{x}^{1} dy\,g(y,Q^{2})
\sigma(\gamma^{*}g\rightarrow q_{f}\bar{q}_{f},yW^{2})\,\,\,,
\label{eq:2.5.1}
\endeq
where $g(y,Q^{2})$ is the distribution function of the
physical, transverse, gluons in the proton. By the kinematics
of DIS
\beq
 y={(kq)\over (pq)} = {(M^{2}+Q^{2})\over (W^{2}+Q^{2})}\,\,.
\label{eq:2.5.2}
\endeq
The diagram of
Fig.~2 describes the driving term of the gluon distribution
in the proton,
which here is assumed to be entirely of the radiative origin
([35,37], see also [39]). At $x\ll 1$ we have [35,37]
\beq
g(x,Q^{2})=
{8 \over x}
\int_{0}^{Q^{2}} { dk^{2} \,k^{2} \over
(k^{2}+\mu_{G}^{2})^{2} }
V(k)
{ \alpha_{S}(\vec{k}\,^{2}) \over 2\pi } \,\,,
\label{eq:2.5.3}
\endeq
Here it is worthwhile to emphasize, that the flux of soft gluons
depends only on the color charge, but neither the spin nor
the helicity, of the source of gluons.
The vertex function $V(\vec{k})$ in the integrand of
(\ref{eq:2.5.3})
is precisely the same as in eq. (\ref{eq:2.1.4}) and describes
the destructive interference of the radiation of gluons by different
quarks bound in the color-singlet nucleon.
In terms of the mass $M$ of the excited $q\bar{q}$ pair,
Eqs.~(\ref{eq:2.5.1}),(\ref{eq:2.5.3}) give
\beq
\Delta\sigma_{tot}(\gamma^{*}N,W^{2},Q^{2}) \propto
\int_{4m_{f}^{2}}^{W^{2}}
{dM^{2}\over Q^{2}+M^{2}}
\sigma(\gamma^{*}g\rightarrow q\bar{q},M^{2})\, .
\label{eq:2.5.4}
\endeq
Because of the spin-${1\over 2}$ exchange in the $t$-channel,
the cross section
of the photon-gluon fusion subprocess decreases at large $M^{2}$,
\beq
\sigma(\gamma^{*}g\rightarrow q\bar{q},M^{2}) \propto
{1\over Q^{2}+M^{2}}\log\left({(Q^{2}+M^{2})\over 4m_{f}^{2}}\right) \, ,
\label{eq:2.5.5}
\endeq
the integral (\ref{eq:2.5.4}) converges at
finite $M^{2}$ and gives the constant photoabsorption cross
section at small $x$.
 For the closely related reason, one finds the rapidly
convergent mass spectrum (\ref{eq:2.4.4}).
On the other hand, if the $q\bar{q}g$ final state
is produced in the photon-gluon fusion, then because of the spin-1
gluon exchange in the t-channel
$\sigma(\gamma^{*}g\rightarrow q\bar{q}g,M^{2}) \propto const$,
which leads to the rising $\propto \log(1/x)$
contribution to the photoabsorption
cross section [41].
Notice, that excitation of the $q\bar{q}g$
final state in the photon-gluon fusion can be reinterepreted
as the scattering on the nucleon of the $q\bar{q}g$ Fock state
of the photon.
Therefore, one has to study the effect of
higher Fock states of the photon.



\section{Interactions of the $q\bar{q}g$ Fock state of the photon}



\subsection{Interaction cross section for the $q\bar{q}g$ state}

One can easily write down the interaction cross section
for the color-singlet three-parton
$q\bar{q}g$ state using only the color gauge invariance
considerations (the separation of partons in the impact-parameter
space is shown in Fig.~4):
\beq
\sigma_{3}(r,R,\rho)={9 \over 8}[\sigma(R)+\sigma(\rho)] -
{1 \over 8}\sigma(r)                         \, \, .
\label{eq:3.1.1}
\endeq
Indeed, when separation of the quark and of the
antiquark is small,
$r\ll \rho\approx R$, then the $q\bar{q}$ pair will be
indistinguishable from the pointlike colour-octet charge.
In this limit
\beq
\sigma_{3}(0,\rho,\rho)={9 \over 4}\sigma(\rho)     \,\, .
\label{eq:3.1.2}
\endeq
where $9/4$ is the familiar ratio of the octet and triplet
strong couplings. In the opposite limiting cases of $R=0$ or
$\rho=0$ the gluon and the (anti)quark with vanishing separation
are indistiguishable from the pointlike (anti)quark and
\beq
\sigma_{3}(r,0,r)=\sigma_{3}(r,r,0)=\sigma(r)    \,\, .
\label{eq:3.1.3}
\endeq
The  formal derivation goes as follows: In the integrand
of the cross section (\ref{eq:2.1.4}),
the two propagators
$1/(\vec{k}^{2}+\mu_{G}^{2})$ correspond to the Fourier transforms
$U(\vec{k}) =\int d^{3}\vec{r}
\exp(-i\vec{k}\vec{r})\exp(-\mu_{G}r)/r$ and
$U(-\vec{k})$
of the (infrared-regulated) gluonic Coulomb potential.
If the color charge is located at the position $\vec{c}$,
then $U(\vec{k},\vec{c})=U(\vec{k})\exp(-i\vec{k}\vec{c})$.
Whenever the two
exchanged gluons couple to the same parton, one gets the
square of the corresponding strong charge. If the gluons
couple to the two partons located at points $\vec{r}_{1}$
and $\vec{r}_{2}$, the corresponding contribution
acquires the extra phase factor
$\exp[i\vec{k}(\vec{r}_{2}-\vec{r}_{1})]$. Precisely
this is the origin of the factor
$[1-\exp(i\vec{k}\vec{r})]$ in the
integrand of eq.~(\ref{eq:2.1.4}). The accurate calculation
of the colour traces for the different couplings of the
two exchanged
gluons to the quark, antiquark and gluon of the
$q\bar{q}g$ Fock state leads  precisely to
the cross section (\ref{eq:3.1.1}).

If  $\Phi_{1}(\vec{r},\vec{R},\vec{\rho},z,z_{g})$ is the
wave function of the $q\bar{q}g$ Fock state, then the
corresponding contribution to $\sigma_{tot}(\gamma^{*}p)$
shall equal
\beq
\Delta\sigma_{tot}(q\bar{q}g,x,Q^{2})=
\int
dz\,d^{2}\vec{r}\,
dz_{g}\, d^{2}\vec{\rho} \,\,
|\Phi_{1}(\vec{r},\vec{R},\vec{\rho},z,z_{g})|^{2}
\sigma(r,R,\rho)\,\, .
\label{eq:3.1.4}
\endeq
The gluon of the $q\bar{q}g$  Fock state is generated radiatively
from the primary $q\bar{q}$  Fock state (Fig.5),
and this radiation
simultaneously renormalizes the weight of the $q\bar{q}$ component
of the photon. If $n_{g}(z,\vec{r})$ is the number of gluons in
the $q\bar{q}g$ state with the $q$-$\bar{q}$ separation $\vec{r}$
(we suppress the subscripts $T$ and $L$ in the $|\Psi(z,r)|^{2}$)
,
defined by
\beq
\int
dz_{g}\,d^{2}\vec{\rho} \,\,
|\Phi_{1}(\vec{r},\vec{R},\vec{\rho},z,z_{g})|^{2}=
n_{g}(z,\vec{r})
|\Psi(z,r)|^{2}                          \, ,.
\label{eq:3.1.5}
\endeq
then the wave function of the
radiationless $q\bar{q}$ component  of the photon shall
renormalize as
\beq
|\Psi_{q\bar{q}}(z,r)|^{2}
=
|\Psi(z,r)|^{2}[1-n_{g}(z,\vec{r})] \,\, .
\label{eq:3.1.6}
\endeq
It is convenient to introduce
\beq
\Delta \sigma(r,R,\rho)=\sigma_{3}(r,R,\rho) -
\sigma(r)=
{9 \over 8}[\sigma(R)+\sigma(\rho)-\sigma(r)]   \, \, ,
\label{eq:3.1.7}
\endeq
which shows how much the interaction cross section of
the $q\bar{q}g$ Fock state is different from the
cross section for the
$q\bar{q}$ state.
Then, $\sigma_{tot}(\gamma^{*}N)$ from interactions of the
$q\bar{q}$ and $q\bar{q}g$ Fock states of the photon takes the form
\arr
\sigma_{tot}(\gamma^{*}N,x,Q^{2})=~~~~~~~~~~~~~~~~~~~~~~~~~~~~
{}~~~~~~~~~~~~~~~~~~\nonumber\\
\int dz d^{2}\vec{r}\,|\Psi(z,r)|^{2}[1-n_{g}(z,\vec{r})]\sigma(r)+
\int dz d^{2}\vec{r}\,dz_{g}\vec{\rho}\,\,
|\Phi_{1}(\vec{r},\vec{R},\vec{\rho},z,z_{g})|^{2}
[\sigma(r)+\Delta\sigma(r,R,\rho)] \nonumber\\
=\int dz d^{2}\vec{r}\,|\Psi(z,r)|^{2}\sigma(r)+
\int dz d^{2}\vec{r}\,dz_{g}\vec{\rho}\,\,
|\Phi_{1}(\vec{r},\vec{R},\vec{\rho},z,z_{g})|^{2}
\Delta\sigma(r,R,\rho)\,\, .  ~~~~~~~~~~~~~~
\label{eq:3.1.8}
\endarr

Up to now we have manipulated the formally divergent
quantity   $n_{g}(z,\vec{r})$ as if it   were finite.
As a matter of fact, the
renormalization (\ref{eq:3.1.6}) of the
radiationless $q\bar{q}$ Fock state
corresponds to the introduction of the so-called
regularized splitting functions [13]
in the GLDAP evolution
equations, and takes care of the virtual radiative
corrections (a very detailed discussion of the
interplay of the virtual and real radiative corrections and
of the emergence of the running strong coupling was given
by Dokshitzer [12],  see also the review paper [40], and needs
not be repeated here).
 Of course, the final result for the physical
cross section, the last line of eq.~(\ref{eq:3.1.8}), does
not contain any divergences.



\subsection
{Wave function of the $q\bar{q}g$ state and the rising
photoabsorption cross section}

We are interested in the $\propto \log(1/x)$ component of the
increase of the photoabsorption cross section
\arr
\Delta \sigma_{tot}^{(1)}(\gamma^{*}N,x,Q^{2})=
\int dz d^{2}\vec{r}\,dz_{g}d^{2}\vec{\rho}\,\,
|\Phi_{1}(\vec{r},\vec{R},\vec{\rho},z,z_{g})|^{2}
\Delta\sigma(r,R,\rho)\,\, .  ~~~~~~~~~~~~~~
\label{eq:3.2.1}
\endarr
This $\log(1/x)$ comes from
the $dz_{g}/z_{g}$ integration in the domain $x < z_{g} <1$ and
we must concentrate on $z_{g} \ll 1$.
One should not confuse the Sudakov variable
$z_{g}$ which is the fraction of
the light-cone momentum of the photon carried by the gluon,
with $y$, which is the fraction of
the light-cone momentum of the proton carried by the same gluon.
The two quantities  are related by
\beq
z_{g}y=x  \, .
\label{eq:3.2.2}
\endeq
Only the diagrams of Fig.~6a in which the exchanged gluons couple
to the gluon of the $q\bar{q}g$ Fock state give rise to
$\Delta \sigma_{tot}(\gamma^{*}N,x,Q^{2}) \propto
\log(1/x)$. The corresponding wave function can be
reconstructed from the number of gluons $n_{g}$ in the
$q\bar{q}g$ state, which on the one hand equals
\beq
n_{g}=\int dz_{g} d^{2}\vec{\rho} dz d^{2}\vec{r} \,\,
|\Phi_{1}(\vec{r},\vec{R},\vec{\rho},z,z_{g})|^{2} \,
\label{eq:3.2.3}
\endeq
and, on the other hand, can be evaluated from the
Weizs\"acker-Williams formula (\ref{eq:2.5.3})
\arr
n_{g}={2 \over 3}\cdot{8 \over\pi}\int {dz_{g}\over z_{g}}
\int dz d^{2}\vec{r}  \,\,
|\Psi(\vec{r},z)|^{2}
\cdot\int { d^{2}\vec{k}_{g}\,\vec{k}_{g}^{2} \over
(\vec{k}^{2}_{g}+\mu_{G}^{2})^{2}} {\alpha_{S}(k_{g}^{2}) \over 2\pi}
\,\,[1-\exp(i\vec{k}_{g}\vec{r})]    \, .
\label{eq:3.2.4}
\endarr
Here we have used the fact that for the $q\bar{q}$ source of
gluons
\beq
V(\vec{k}_{g})=\int dz d^{2}\vec{r} \,\,
|\Psi(\vec{r},z)|^{2}[1-\exp(i\vec{k}_{g}\vec{r})]   \, ,
\label{eq:3.2.5}
\endeq
and the factor $2/3$ accounts for the 2 (anti)quarks in the
$q\bar{q}$ state compared to the 3 quarks in the proton.
Transformation of Eq.~(\ref{eq:3.2.4}) into the configuration-space
integral can easily be performed making use of [24]
\beq
\int { d^{2}\vec{k}\,\, \vec{k}^{2} \over
(\vec{k}^{2}+\mu_{G}^{2})^{2}}
=\int d^{2}\vec{\rho}\,\mu_{G}^{2}K_{1}(\mu_{G}\rho)^{2}
\label{eq:3.2.6}
\endeq
and
\beq
\int { d^{2}\vec{k}\,\, \vec{k}^{2} \over
(\vec{k}^{2}+\mu_{G}^{2})^{2}} \exp(i\vec{k}\vec{r})
=\int d^{2}\vec{\rho}\,\mu_{G}^{2}
K_{1}(\mu_{G}\rho)K_{1}(\mu_{G}R){\vec{R}\vec{\rho}\over R\rho}
     \, ,
\label{eq:3.2.7}
\endeq
which yields
\arr
n_{g}={4 \over 3\pi^{2}}\int {dz_{g}\over z_{g}}\,d^{2}\vec{\rho}
\int dz d^{2}\vec{r} \,\,
|\Psi(\vec{r},z)|^{2}\alpha_{S}(r)\mu_{G}^{2}
|K_{1}(\mu_{G}\rho){\vec{\rho}\over \rho}
-K_{1}(\mu_{G}R){\vec{R}\over R}|^{2}     \,\, ,
\label{eq:3.2.8}
\endarr
so that
\arr
|\Phi_{1}(\vec{r},\vec{R},\vec{\rho},z,z_{g})|^{2}=
{1 \over z_{g}} {4 \over 3\pi^{2}}
|\Psi(z,r)|^{2}\alpha_{S}(r)\mu_{G}^{2}
|K_{1}(\mu_{G}\rho){\vec{\rho}\over \rho}
-K_{1}(\mu_{G}R){\vec{R}\over R}|^{2} \, .
\label{eq:3.2.9}
\endarr
The wave function (\ref{eq:3.2.9}) has the $1/z_{g}$ behavior
needed for the $\propto \log(1/x)$ rise of the cross section.
The color gauge invariance property of the wave function
(\ref{eq:3.2.9}) is noteworthy: because of cancellations
of the color charges of the quark and antiquark in the
color singlet state it vanishes when $(R-\rho) \rightarrow 0$.
It counts only physical, transverse gluons.
For those reasons  and because of the related color gauge
invariance properties of $\sigma(r,R,\rho)$, introduction of
the infared regularization and the modelling of the confinement
by the effective mass of gluons exchanged in the $t$-channel
does not conflict the color gauge invariance.

In the DIS the leading contribution to
$\Delta \sigma_{tot}^{(1)}(\gamma^{*}N,x,Q^{2})$
comes from the LLA ordering of
sizes
\beq
{1 \over Q^{2}} \ll r^{2} \ll \rho^{2}\ll
R_{N}^{2} ,{1\over \mu_{G}^{2}} \,\, ,
\label{eq:3.2.10}
\endeq
when (here the
angular averaging is understood)
\beq
\mu_{G}^{2}|K_{1}(\mu_{G}\rho){\vec{\rho} \over \rho}
-K_{1}(\mu_{G}R){\vec{R}\over R}|^{2}=
{r^{2} \over \rho^{4}} \, ,
\label{eq:3.2.11}
\endeq
which produces the factorized wave function
\beq
|\Phi_{1}(r,R,\rho)|^{2}={1\over z_{g}}|\Psi(z,r)|^{2}
{4 \over 3\pi^{2}}\alpha_{S}(r){r^{2} \over \rho^{4}} \, .
\label{eq:3.2.12}
\endeq
Naturally, the LLA wave function (\ref{eq:3.2.12}) does not
depend on the infrared regularization parameter $\mu_{G}$.
In the LLA
\beq
  \Delta\sigma(r,R,\rho)= \Sigma(\rho)=
{9 \over 4} \sigma(\rho)
\label{eq:3.2.13}
\endeq
and the increase of the total cross section can be written as
\arr
\Delta \sigma_{tot}(\gamma^{*}N,x,Q^{2})=
\int dz\,d^{2}\vec{r}\,
|\Psi(z,r)|^{2}\alpha_{s}(r)r^{2}
 {4 \over 3\pi}
\int_{x}^{z} {dz_{g}\over z_{g}}
\int_{r^{2}} {d\rho^{2} \over \rho^{4}}
\Sigma(\rho) ~~~~
 \nonumber \\
=
\int dz\,d^{2}\vec{r}\,
|\Psi(z,r)|^{2}\alpha_{s}(r)C_{N}r^{2}
{9 \over 4} {4 \over 3\pi}
\log\left({z\over x}\right)
\int_{r^{2}} {d\rho^{2} \over \rho^{4}}
\rho^{2}\alpha_{S}(\rho)L(\rho) ~~~~~~~~~~
  \nonumber \\
=\int dz\,d^{2}\vec{r}\,
|\Psi(z,r)|^{2}\sigma(r)
{12\over \beta_{0}}\cdot
 {1\over 2!}L(r)\cdot {1 \over 1!}
 \log\left({z\over x}\right)
 \propto {1\over Q^{2}}L(Q)^{3}\log\left({1\over x}\right)\,\, .
\label{eq:3.2.14}
\endarr
Here we have made an explicit use of the small-$r$ behaviour
of $\sigma(r)$, Eq.~(\ref{eq:2.2.1}).
 This is the first instance when we encounter
 the expansion parameter of the
 {\sl Double-Leading-Logarithm
Approximation}  (DLLA) [12,40,42]
\beq
\xi(x,r)={12\over \beta_{0}}L(r)\log\left({1\over x}\right) \, .
\label{eq:3.2.15}
\endeq
We have one power of $L(Q)$ per
QCD loop (which {\sl \`{a} posteriori} justifies LLA ordering
(\ref{eq:3.2.10})) and one power of $\log(1/x)$ per gluon in
the Fock state of the photon.



\subsection{The triple-pomeron asymptotics of the mass spectrum
of diffraction dissociation}

Our starting point is the generic
formula (\ref{eq:1.7}). Repeating the considerations of
Section 3.2, we can write
\arr
16\pi\left.{d\sigma_{D}(\gamma^{*}\rightarrow
q\bar{q} +q\bar{q}g)
 \over dt}\right|_{t=0}=
16\pi\int dM^{2}\,
\left.{d\sigma_{D}(\gamma^{*}\rightarrow q\bar{q}+q\bar{q}g)
\over dt dM^{2}}\right|_{t=0}=
{}~~~~~~~~~~\nonumber \\
\int dz d^{2}\vec{r}\,\,
|\Psi(z,r)|^{2}[1-n_{g}(\vec{r})]\sigma(r)^{2}+
\int dz d^{2}\vec{r}dz_{g}d^{2}\vec{\rho}\,\,
|\Phi_{1}(\vec{r},\vec{R},\vec{\rho},z,z_{g})|^{2}
[\sigma(r)+\Delta\sigma(r,R,\rho)]^{2}= \nonumber\\
\int dz d^{2}\vec{r}\,\,|\Psi(z,r)|^{2}\sigma(r)^{2}+
\int dz d^{2}\vec{r}dz_{g}d^{2}\vec{\rho}\,\,
|\Phi_{1}(\vec{r},\vec{R},\vec{\rho},z,z_{g})|^{2}
[\Delta\sigma(r,R,\rho)^{2}+
2\sigma(r)\Delta\sigma(r,R,\rho)]\, .~~~
\label{eq:3.3.1}
\endarr
The first term in the last line of Eq.~(\ref{eq:3.3.1})
describes the diffraction excitation of
the $q\bar{q}$ Fock states into the
low masses $M^{2}\sim Q^{2}$, see Eq.~(\ref{eq:2.4.4}).
The second term gives rise to the $1/M^{2}$ mass spectrum,
which can be seen as follows:
The invariant mass squared of the $q\bar{q}g$ state equals
\beq
M^{2}=
{m_{f}^{2}+k_{q}^{2} \over z} +
{m_{f}^{2}+k_{\bar{q}}^{2} \over 1-z-z_{g}}   +
{k_{g}^{2} \over z_{g}}
\label{eq:3.3.2}
\endeq
Anticipating the final
results, we note that the leading contribution
to the diffraction dissociation cross section comes
from the slightly modified LLA ordering
\beq
{1\over Q^{2}}\ll r^{2} \ll \rho^{2} \sim
R_{N}^{2},\,{1\over \mu_{G}^{2}} \, ,
\label{eq:3.3.3}
\endeq
{\sl i.e.}, from $Q^{2} \gg k_{q}^{2},k_{\bar{q}}^{2}
\gg k_{g}^{2} \sim \mu_{G}^{2}$. Therefore, the excitation of
masses $M^{2} \gg Q^{2}$ only comes from $z_{g} \ll z<1$, and the
$dz_{g}$ integration in (\ref{eq:3.3.1}) can easily be
transformed into the $dM^{2}$ integration
(see Eqs.~(\ref{eq:2.5.2},\ref{eq:3.2.2})):
\beq
{dM^{2} \over M^{2}+Q^{2} } = {dy\over y}={dz_{g} \over z_{g} }  \, ,
\label{eq:3.3.4}
\endeq
where now $y$ is the fraction of proton's momentum carried by
the pomeron.
The wave function (\ref{eq:3.2.9}) has precisely the needed
$\propto 1/z_{g}$ behaviour
and leads to
(in view of (\ref{eq:3.3.3}) the term
$\propto \sigma(r)\Sigma(r,R,\rho)$ in (\ref{eq:3.3.1}) can be
neglected)
\arr
\left.{d\sigma_{D} \over dt dM^{2}}\right|_{t=0}
={1\over M^{2}+Q^{2}}\int dz \,d^{2}\vec{r}\,\,
|\Psi(z,r)|^{2}\alpha_{S}(r) r^{2} C_{N}\nonumber\\
\cdot{1\over C_{N}}
\cdot{1 \over 16\pi}\cdot{4 \over 3\pi}
\cdot \int_{r^{2}}^{\infty}d\rho^{2}
\left[{\Sigma(\rho)\over \rho^{2}}\right]^{2} F(\mu_{G}\rho) \,,
\label{eq:3.3.5}
\endarr
where $F(\mu_{G}\rho)$ is defined by the slight generalization
of (\ref{eq:3.2.11}) to allow for $\mu_{G}\rho \sim 1$:
\beq
\mu_{G}^{2}|K_{1}(\mu_{G}\rho){\vec{\rho}\over\rho}
-K_{1}(\mu_{G}R){\vec{R}\over R}|^{2} =
{r^{2}\over \rho^{4}}F(\mu_{G}\rho) \, .
\label{eq:3.3.6}
\endeq
The form factor $F(x)$ satisfies $F(0)=1$ and
$F(x) \propto\exp(-2x)$ at $x > 1$.

Firstly, we notice that
the diffraction dissociation cross section depends on
the infrared regularization, since the
$\rho$ integration in (\ref{eq:3.3.5}) is essentially flat
and is dominated by large $\rho \sim R_{N},1/\mu_{G}$. Then
can take $\rho^{2}=0$ for the lower limit of intergation,
and  $d\sigma_{D}/dtdM^{2}$ Eq.~(\ref{eq:3.3.5}) factorizes
into the $Q^{2}$-independent dimensional constant
\beq
A_{3\Pom}^{*}={1 \over C_{N}}\cdot {1\over 12\pi^{2}}\cdot
\left({9 \over 4}\right)^{2}
\int d\rho^{2}\,\,
\left[{\sigma(\rho)\over \rho^{2}}\right]^{2} F(\mu_{G}\rho)
\label{eq:3.3.7}
\endeq
and the cross section
\beq
\sigma^{*}(Q^{2})=
\int dz \,d^{2}\vec{r}\,|\Psi(z,r)|^{2}C_{N}r^{2}\alpha_{S}(r)\,\, ,
\label{eq:3.3.8}
\endeq
which is nearly identical to $\sigma_{T}(\gamma^{*}N,x,Q^{2})$
Eq.~(\ref{eq:2.3.1}), being short of $L(r)$ in the integrand.
Therefore, this driving term of the triple-pomeron component
of the diffraction dissociation of virtual photons satisfies
the approximate factorization reminiscent of factorization
properties of
the triple-pomeron
diagram of the conventional Regge theory Fig.~7a,
\beq
{M^{2}+Q^{2} \over \sigma_{T}(\gamma^{*}N,x,Q^{2})}\cdot
\left.{d\sigma_{D}(\gamma^{*}+N\rightarrow X+N) \over
dtdM^{2}}\right|_{t=0} = A_{3\Pom}(Q^{2})=
{\sigma^{*}(Q^{2}) \over \sigma_{T}(\gamma^{*}N,x,Q^{2})}
A^{*}_{3\Pom}
\label{eq:3.3.9}
\endeq
To the considered lowest order in the
perturbation theory, the quantity $A_{3\Pom}(Q^{2})$ does not depend
on $x$. $A_{3\Pom}(Q^{2})$ is the dimensionfull quantity, and as
such it could have had a strong $Q^{2}$-dependence,
$A_{3\Pom}(Q^{2}) \sim 1/Q^{2}$ being a plausible guess if
$1/\sqrt{Q^{2}}$ were the only  scale relevant to deep inelastic
scatering. The fact that $\sigma^{*}(Q^{2})$ and
$\sigma_{tot}(\gamma^{*}N,x,Q^{2})$ are nearly identical, proves
that this is not the case. Furthermore,
the {\sl r.h.s} of Eq.~(\ref{eq:3.3.9}) has a very smooth
extrapolation down to the real photoproduction limit $Q^{2}=0$,
confirming the earlier suggestions [26,7] that
$A_{3\Pom}(Q^{2})$ is close to $A_{3\Pom}(0)$ as measured in
the real photoproduction (the more detailed comparison of
$A_{3\Pom}(Q^{2})$ with the $A_{3\Pom}$ for the diffraction dissociation
of protons, pions and the real photons will be presented elsewhere [43]).
For the order of magnitude estimation
of $A_{3\Pom}^{*}$,
in the dominant region of integration in (\ref{eq:3.3.7})
we can take
${\cal F}(\mu_{G}\rho)\sim \exp(-2\mu_{G}\rho)$,
$L(\rho)\sim 1$ and
$\sigma(\rho)/\rho^{2} \sim C_{N}
\alpha_{S}(\rho={1\over 2\mu_{G}})$
with the result
\beq
A_{3\Pom}^{*} \sim {27C_{N} \over 256\pi^{2}}
{\alpha_{S}(4\mu_{G}^{2})^{2}\over
\mu_{G}^{2}} =
{\alpha_{S}(4\mu_{G}^{2})^{2}\over
8\mu_{G}^{2}}  \, .
\label{eq:3.4.10}
\endeq
With $\mu_{G} \sim 0.4\,GeV $ this gives $A_{3\Pom}^{*} \sim
0.3\,(GeV)^{-2}$. The experimental data on the diffraction
dissociation
of the real photons give
$A_{3\Pom}(0)\approx 0.16\, (GeV)^{-2}$ [27].
This dimensional soupling $A_{3\Pom}^{*} \approx A_{3\Pom}(0)$
emerges as the principal nortmalization factor of the
diffraction dissociation cross section, and Eq.~(\ref{eq:3.3.9})
is a starting point of the derivation of the factorization
representation (\ref{eq:6.3}) to all orders in the perturbation
theory.

Combining Eq.~(\ref{eq:3.3.9}) with the definition
Eq.~(\ref{eq:2.4.5}) we find the corresponding contribution to
the structure function of the pomeron at $x=Q^{2}/(Q^{2}+M^{2})\ll 1$
of the form [7]
\beq
F_{2}^{(\Pom)}(x,Q^{2})={4Q^{2} \over
\pi\alpha_{em}\sigma_{tot}(pp)}
\sigma^{*}(x,Q^{2})A_{3\Pom}^{*}
\approx {16\pi A_{3\Pom}(0) \over \sigma_{tot}(pp)}
F_{2}^{(N)}(x,Q^{2}) \approx 0.08F_{2}^{(N)}(x,Q^{2}) \, \, .
\label{eq:3.3.11}
\endeq
It describes DIS on the $q\bar{q}$ 'sea' of the pomeron.
The relationship (\ref{eq:3.3.11}) shows a deep connection between
the triple-pomeron component of the mass spectrum and the sea
structure function of the pomeron.
Notice the difference between
the diffractive excitation of the $q\bar{q}$
state, Fig.~3, and of the $q\bar{q}g$ state, Fig.~7b : in the former
the pomeron couples to (anti)quarks and the DIS probes the 'valence'
$q\bar{q}$ structure of the pomeron, in the latter the pomeron
couples to gluons, and the DIS probes the 'sea' of the pomeron,
which can be treated as having been
generated from the 'valence' (constituent) gluons of
the pomeron. The diffraction dissociation cross section
(\ref{eq:3.3.5}) and the pomeron structure function (\ref{eq:3.3.11})
are sensitive to the infrared regularization, and the normalization
of both quantities contains the new dimensional parameter
$A_{3\Pom}^{*}$. The important result
of the above analysis is that this
dimensional parameter $A_{3\Pom}^{*}$
can approximately be inferred from the real photoproduction data.
Notice a close similarity between Eq.~(\ref{eq:3.3.7})
for the normalization of the triple-pomeron mass spectrum
and Eq.~(\ref{eq:2.4.1}) for the normalization of the mass spectrum
for the excitation of the $q\bar{q}$ state. However, whereas
Eqs.~(\ref{eq:3.3.7},\ref{eq:3.3.11}) predict the flavor-independent
relation between the proton and pomeron structure functions, the
valence $q\bar{q}$ structure function of the pomeron has a strong
flavor dependence [7].
Now we shall study how these conclusions shall change when higher order
effects and QCD evolution are included.


\section{Rising structure functions and higher order Fock
states of the photon}

Generalization of an analysis of Section 3
to interactions of
the higher $q\bar{q}g_{1}...g_{n}$ Fock states
of the photon is straightforward.
The strong DLLA ordering of gluons
\beq
x\ll z_{n} \ll z_{n-1} \ll ...\ll z_{1} \ll z<1  \, ,
\label{eq:4.1}
\endeq
\beq
{1 \over Q^{2}} \ll r^{2} \ll \rho_{1}^{2}\ll ....\ll
\rho_{n}^{2} \ll R_{c}^{2}   \, .
\label{eq:4.2}
\endeq
is required to have maximal possible powers of $\log(1/x)$ and
$L(Q)$ [42,12]. To the DLLA
the quark-loop insertions in the generalized ladder diagrams
of Fig.~8 can be neglected. By virtue of the ordering of sizes
(\ref{eq:4.2}) the $q\bar{q}g_{1}...g_{n}$ Fock
state interacts like the colour-singlet octet-octet state,
with the inner subsystem $q\bar{q}g_{1}...g_{n-1}$ acting
like the pointlike colour-octet charge. Henceforth, the
generalization of Eq.~(\ref{eq:3.2.13}) is
\beq
\Delta \sigma(r,\rho_{1},....,\rho_{n}) = \Sigma(\rho_{n})
={9 \over 4}\sigma(\rho_{n})  \, .
\label{eq:4.3}
\endeq

The DLLA  wave function is a straightforward
generalization of the wave function (\ref{eq:3.2.12})
for the $q\bar{q}g$ Fock state:
\arr
|\Psi(r,\rho_{1},...,\rho_{n})|^{2}= ~~~~~~~~~~~~~~~~~~~~~~~
{}~~~~~~~~\nonumber\\
|\Psi(z,r)|^{2}\cdot
\cdot {1\over z_{1}}\alpha_{S}(r){4\over 3\pi^{2}}
{r^{2}\over \rho_{1}^{4}}
\cdot {1\over z_{2}}\alpha_{S}(\rho_{1}){3\over \pi^{2}}
{\rho_{1}^{2}\over \rho_{2}^{4}}\cdot ... \cdot
{1\over z_{n}}
{3 \over \pi^{2}}\alpha_{S}(\rho_{n-1})
{\rho_{n-1}^{2} \over \rho_{n}^{4}} \,\,  .
\label{eq:4.4}
\endarr
Notice, that the first gluon is radiated by the
triplet-antitriplet colour-singlet state. The subsequent
gluons are radiated by the octet-octet colour-singlet
state, which brings in the ratio $9/4$ of the octet and
triplet strong couplings.
The corresponding increase of the total cross section equals
(here we make an explicit use of eq.~(\ref{eq:2.2.1}))
\arr
\Delta\sigma_{tot}^{(n)}(\gamma^{*}N,x,Q^{2})=
{4\over 3\pi}\cdot
 \left({3\over \pi}\right)^{n-1}
\int dz\, d^{2}\vec{r}\,|\Psi(z,r)|^{2}
\alpha_{S}(r)r^{2}~~~~~~~~~~~~~~~~~~~~~~~
\nonumber\\
\cdot\int_{r^{2}} {d\rho_{1}^{2} \over \rho_{1}^{2}}
\alpha_{S}(\rho_{1})
\int_{\rho_{1}^{2}} {d\rho_{2}^{2} \over \rho_{2}^{2}}
\alpha_{S}(\rho_{2})...
\int_{\rho_{n-1}^{2}} {d\rho_{n}^{2} \over \rho_{n}^{4}}
\Sigma(\rho_{n})\cdot
\int_{x}^{z}{dz_{1}\over z_{1}}
\int_{x}^{z_{1}}{dz_{2}\over z_{2}} ...
\int_{x}^{z_{n-1}}{dz_{n}\over z_{n}} \nonumber \\
=\left({3\over \pi}\right)^{n}
\int dz\, d^{2}\vec{r}\,|\Psi(z,r)|^{2}
\alpha_{S}(r)C_{N}r^{2}~~~~~~~~~~~~~~~~~~~~~~~
\nonumber\\
\cdot\int_{r^{2}} {d\rho_{1}^{2} \over \rho_{1}^{2}}
\alpha_{S}(\rho_{1})
\int_{\rho_{1}^{2}} {d\rho_{2}^{2} \over \rho_{2}^{2}}
\alpha_{S}(\rho_{2})...
\int_{\rho_{n-1}^{2}} {d\rho_{n}^{2} \over \rho_{n}^{2}}
\alpha_{S}(\rho_{n}) L(\rho_{n})\cdot
\int_{x}^{z}{dz_{1}\over z_{1}}
\int_{x}^{z_{1}}{dz_{2}\over z_{2}} ...
\int_{x}^{z_{n-1}}{dz_{n}\over z_{n}} \nonumber \\
=\int dz\, d^{2}\vec{r}\,|\Psi(z,r)|^{2}
\alpha_{S}(r)C_{N}r^{2}
\left({12\over \beta_{0}}\right)^{n}
{1 \over (n+1)!}L(r)^{n+1}
{1\over n!}\log\left({z\over x}\right)^{n} ~~~~~~~~~~~~
\nonumber \\
=
\int dz\, d^{2}\vec{r}\,|\Psi(z,r)|^{2}
\sigma(r)
{\xi(x/z,r)^{n} \over (n+1)!n!} \,\,.~~~~~~~~~~~~~~~~~~~~~~~~~~~
\label{eq:4.5}
\endarr
Therefore, the total photoabsorption cross section can
be represented as
\beq
\sigma_{tot}(x,Q^{2})=
\int dz\, d^{2}\vec{r}\,|\Psi(z,r)|^{2} \sigma({x\over z},r)   \, \,.
\label{eq:4.6}
\endeq
where
\beq
\sigma(x,r)=\sigma(r)
\sum_{n=0}
{\xi(x,r)^{n} \over (n+1)! n!}  \, ,
\label{eq:4.7}
\endeq
is the energy-dependent dipole cross section, which
generalizes the Low-Nussinov pomeron
to the DLLA pomeron.

The sum in (\ref{eq:4.7}) can be evaluated as
(we neglect the slowly varying pre-exponential factor)
\beq
\sum_{n=0}{\xi^{n} \over (n+1)! n!} =
{\partial \over \partial \xi}
\sum_{n=1}{\xi^{n} \over (n!)^{2}} \propto
\exp(2\sqrt{\xi}) \, ,
\label{eq:4.8}
\endeq
leading to
\beq
\sigma(x,r)\propto \sigma(r)\exp\sqrt{{48\over\beta_{0}}
\log\left[{1\over \alpha_{S}(r)}\right]\log\left({1\over x}\right)}
\label{eq:4.9}
\endeq
and
\beq
F_{2}(GLDAP,x,Q^{2})
={Q^{2} \over 4\pi \alpha_{em}}\sigma_{tot}(x,Q^{2})
\propto
\exp\sqrt{{48\over \beta_{0}}
\log\left[{1\over \alpha_{S}(Q^{2})}\right]
\log\left({1\over x}\right)} \, .
\label{eq:4.10}
\endeq

The representation (\ref{eq:4.6}) for the photoabsorption
cross section in terms of the DLLA dipole cross section
(\ref{eq:4.7}) is a new result and is presented here for the first time.
However, since the perturbative expansion (\ref{eq:4.5})
is completely equivalent to the GLDAP evolution equation
for the structure function [11,12], the result
(\ref{eq:4.10}) for the DLLA growth of the structure function
is identical to the one derived from the GLDAP
evolution equations
[12,40], where it appeard as a rising density
$g(GLDAP,x,Q^{2})$
of gluons in the proton. In our light-cone $s$-channel
approach it comes from
interactions of the higher $q\bar{q}g_{1}...g_{n}$ Fock states
of the photon and can be described in terms of the
rising DLLA pomeron cross section (\ref{eq:4.7})
for the color dipole. As a matter of fact, it can easily be shown
that in the DLLA [30]
\beq
\sigma(x,r)= {\pi^{2} \over 3}
r^{2}\alpha_{S}(r)xg(x,Q^{2}={1\over r^{2}})
\label{eq:4.11}
\endeq
We note in passing, that the cross section
(\ref{eq:4.7},\ref{eq:4.9},\ref{eq:4.11}) obviously
does not satisfy the factorization relations usually assumed for
the pomeron in the standard Regge phenomenology. The QCD evolution
analysis of deep inelastic scattering needs not to assume any
factorization of $F_{2}(x,Q^{2})$.


\section{The structure function of the pomeron and constituent
gluons of the pomeron}

Now we consider diffraction excitation of the higher order
$q\bar{q}g_{1}....g_{n+2}$ Fock states of the photon (Fig.~7c).
Large masses $M$ of the excited state
\beq
M^{2} =
{m_{f}^{2}+k_{q}^{2} \over z_{q}} +
{m_{f}^{2}+k_{\bar{q}}^{2} \over z_{\bar{q}}}+
\sum_{i=1}^{n+2}{k_{i}^{2} \over z_{i}}
\label{eq:5.1}
\endeq
will be dominated by the softest gluon contribution.
The fraction $y$ of proton's
momentum carried away by the pomeron
is related to $z_{n+2}$ as (cf. Eq.~(\ref{eq:3.2.2}))
\beq
z_{n+2}= {x\over y}
\label{eq:5.2}
\endeq
and
\beq
{dM^{2} \over M^{2}+Q^{2}} = {dy \over y} =
{dz_{n+2} \over z_{n+2}}  \,\, .
\label{eq:5.3}
\endeq
Using the wave function (\ref{eq:4.4}), and repeating the
considerations which have lead to eq.~(\ref{eq:3.3.5}),
we find the contribution
of excitation of the $q\bar{q}g_{1}...g_{n+2}$ Fock state to
the mass spectrum of the diffraction dissociation and to the
photon-pomeron cross section
\arr
\Delta \sigma_{tot}^{(n+2)}(\gamma^{*}\Pom,Q^{2},M^{2})=
{16\pi \over \sigma_{tot}(pp)}
(M^{2}+Q^{2})\left. { d\sigma_{D}(\gamma^{*}
\rightarrow q\bar{q}g_{1}...g_{n+2})
 \over dt dM^{2}}\right|_{t=0} ~~~~~~~~~~~~~~~~~\nonumber\\
=
 {1 \over \sigma_{tot}(pp)}\cdot\left({4\over 3\pi}\right)\cdot
\left({3\over\pi}\right)^{n+1}
\int dz\, d^{2}\vec{r}\,|\Psi(z,r)|^{2}
\alpha_{S}(r)r^{2}~~~~~~~~~~~~~~~~~~~\nonumber\\
\cdot\int_{r^{2}} {d\rho_{1}^{2} \over \rho_{1}^{2}}
\alpha_{S}(\rho_{1}) ...
\int_{\rho_{n-1}^{2}} {d\rho_{n}^{2} \over \rho_{n}^{2}}\cdot
\alpha_{S}(\rho_{n})
\int_{\rho_{n}^{2}} {d\rho_{n+1}^{2} \over \rho_{n+1}^{2}}
\alpha_{S}(\rho_{n+1})
\int_{\rho_{n+1}^{2}} {d\rho_{n+2}^{2}  \over \rho_{n+2}^{4}}
\Sigma(\rho_{n+2})^{2} {\cal F}(\mu_{G}\rho_{n+2})~~~~
 \nonumber\\
\cdot\int_{x}^{z}{dz_{1}\over z_{1}}
\int_{x}^{z_{1}}{dz_{2}\over z_{2}} ...
\int_{x}^{z_{n}}{dz_{n+1}\over z_{n+1}}  ~~~~~~~~~~~~~~~~~~~~
{}~~~~~~~~~~~~~~~~~~~~~\nonumber\\
=
 {1 \over \sigma_{tot}(pp)}
\left({3\over \pi}\right)^{n}
\int dz\, d^{2}\vec{r}\,|\Psi(z,r)|^{2}
\alpha_{S}(r)r^{2}
\cdot\int_{x}^{1}{dz_{z}\over z_{1}} ...
\int_{x}^{z_{n-1}}{dz_{n}\over z_{n}}
\cdot\int_{r^{2}} {d\rho_{1}^{2} \over \rho_{1}^{2}}
\alpha_{S}(\rho_{1}) ...
\int_{\rho_{n-1}^{2}} {d\rho_{n}^{2} \over \rho_{n}^{4}}
 \nonumber \\
 \cdot
\rho_{n}^{2}\alpha_{S}(\rho_{n})
 \cdot{81\over 4\pi^{2}}\cdot
\int_{\rho_{n}^{2}} {d\rho_{n+1}^{2} \over \rho_{n+1}^{2}}
\alpha_{S}(\rho_{n+1})
\cdot\int_{x}^{z_{n}}{dz_{n+1}\over z_{n+1}}
\int_{\rho_{n+1}^{2}} {d^{2}\vec{\rho}_{n+2}\over \pi\rho_{n+2}^{4}}
\sigma(\rho_{n+2})^{2} {\cal F}(\mu_{G}\rho_{n+2})\,\, .~~~~
\label{eq:5.4}
\endarr
Comparison with Eqs.~(\ref{eq:2.1.4}),(\ref{eq:4.5}) shows that
the last line of Eq.~(\ref{eq:5.4}) can be reinterpreted as
the dipole cross section for interaction with the pomeron
treated as the two-gluon state with the wave function
\beq
|\Psi_{\Pom}(z_{g},\vec{r})|^{2} = {81 \over 8 \pi^{4}}
\cdot {1\over z_{g}}\cdot{1\over \sigma_{tot}(pp)}
\left[{\sigma(r)\over r^{2}}\right]^{2}
{\cal F}(\mu_{G}r) \, ,
\label{eq:5.5}
\endeq
where
\beq
z_{g} = {z_{n+2}\over z_{n+1}}
\label{eq:5.6}
\endeq
is the fraction of pomeron's momentum carried by the gluon.

Indeed, making use of Eq.~(\ref{eq:3.2.5}) for the vertex
function of the two-body system, the dipole cross section
$\sigma_{2g}(\rho)$
for the scattering on the gluon-gluon state can be written
as
\beq
\sigma_{2g}(\rho) =
2\pi
\rho^{2}\,\alpha_{S}(\rho)
\int dz_{g} \int d^{2}\vec{r}\,
|\Psi_{2g}(z,r)|^{2}
\int_{0}^{1/\rho^{2}}
{ dk^{2}\, k^{2}
\over
(k^{2} + \mu_{G}^{2})^{2} }
\alpha_{S}(k^{2})
[1-\exp(i\vec{k}\vec{r})]
\label{eq:5.7}
\endeq
The factor $3/2$ difference from
Eqs.~(\ref{eq:2.1.4},\ref{eq:2.2.1}) is due
to the ratio $9/4$ of the gluon (octet) and the quark (triplet)
strong couplings and the ratio $2/3$
of the number of the constituents gluons in the pomeron
and the number of the constituent quarks in the proton. The
series of transformations of the integrand of (\ref{eq:5.7}),
\arr
\sigma_{2g}(\rho)=
2\pi
\rho^{2}\,\alpha_{S}(\rho)
\int dz_{g} \int d^{2}\vec{r}\,
|\Psi_{2g}(z_{g},r)|^{2}
\int_{1/r^{2}}^{1/\rho^{2}}
{ dk^{2}
\over
k^{2} }
\alpha_{S}(k^{2}) ~~~~~~~~
\nonumber \\
 =
2\pi
\rho^{2}\,\alpha_{S}(\rho)
\int dz_{g} \int d^{2}\vec{r}\,
|\Psi_{2g}(z_{g},r)|^{2}
\int^{r^{2}}_{\rho^{2}}
{ d\rho_{1}^{2}
\over
\rho_{1}^{2} }
\alpha_{S}(\rho_{1}) ~~~~~~~~~~~~~~~
\nonumber \\
=
2\pi \rho^{2}\,\alpha_{S}(\rho)
\int_{\rho^{2}} {d\rho_{1}^{2} \over \rho_{1}^{2}}
\alpha_{S}(\rho_{1})
\int dz_{g} \int _{\rho_{1}^{2}}d^{2}\vec{r} \,
|\Psi_{2g}(z_{g},r)|^{2} \,\,,~~~~~
\label{eq:5.8}
\endarr
and comparison with the last line of Eq.~(\ref{eq:5.4})
complete  the derivation of the wave function (\ref{eq:5.5}).
This is one of the central results of the present paper.

The wave function $\Psi_{2g}(z_{g},r)$ gives the distribution
of the 'valence' gluons $g_{\Pom}(z_{g})$ in the pomeron
\beq
g_{\Pom}(z_{G})=\int d^{2}\vec{r}\,|\Psi_{2g}(z_{g},r)|^{2}\,\, .
\label{eq:5.9}
\endeq
The perturbative expansion (\ref{eq:5.4}) describes the
QCD evolution of this 'valence' gluon distribution.
The above derivation holds at
$z_{g}\ll 1$ , the
region of $z_{g} \sim 1 $ requires special consideration.
Only $z_{g}\ll 1$ is important in the DLLA.
 The radius of the
pomeron $R_{\Pom} \sim R_{N},1/\mu_{G}$ and is controlled
by both the form factor ${\cal F}(\mu_{G}r)$ and the
behavior of $\sigma(r)$ in the saturation regime.
The absolute normalization of the flux of soft gluons in the
pomeron is given by the familiar coupling $A_{3\Pom}^{*}$:
\beq
z_{g}g_{\Pom}(z_{g}) =
{81 \over 8 \pi^{3}}\cdot
{1\over \sigma_{tot}(pp)}
\int dr^{2}\,\,
\left[{\sigma(r)\over r^{2}}\right]^{2}
{\cal F}(\mu_{G}r)=
{128 \pi \over 9}\cdot {A_{3\Pom}^{*}
\over \sigma_{tot}(pp)} \sim 0.06 \,\, .
\label{eq:5.10}
\endeq
Extrapolation of (\ref{eq:5.10}) also to large $z_{g}$
gives an estimate of the gluon momentum integral for the pomeron
\beq
\langle x_{g} \rangle_{\Pom}=
\int_{0}^{1} dx \,xg_{\Pom}(x) \approx
{128\pi  \over 9}\cdot {A_{3\Pom}^{*}
\over \sigma_{tot}(pp)} \sim 0.06 \,\, .
\label{eq:5.11}
\endeq
The momentum integral for the 'valence' (anti)quarks of the
pomeron, discussed in Section 2.4, was estimated in Ref.~[7]
with the result $\langle x_{v}(q\bar{q}) \rangle_{\Pom} \sim 0.1$.
Eq.~(\ref{eq:3.3.11}) gives a few per cent estimate for the
momentum integral for the sea (anti)quarks.
Here we just emphasize that the pomeron
needs not be regarded as a particle and
on the purely theoretical
grounds there are no reasons why the momentum integrals
for gluons and (anti)quarks in the pomeron
must add to $100\%$ [5,7].

Henceforth, we have identified the three
components of the input for the QCD evolution of the pomeron
structure function: i) the valence quark-antiquark component
with the structure function (\ref{eq:2.4.5}) (for the detailed
analysis see Ref.~[7]), ii) the valence gluon distribution
with the structure function (\ref{eq:5.10}), iii) the sea
(anti)quark distribution given by Eq.~(\ref{eq:3.3.11}).
All these input parton distributions are sensitive to the
infrared regularization. There is nothing wrong with this
sensitivity: the infrared sensitivity of the parton distributions
is inherent to the QCD-improved parton model. In the conventional
parton model phenomenology it is hidden in the parametrization
of the parton densities at small factorization scale,
which then is used as an input in the QCD evolution analysis of
the scaling violations. The important finding is that the absolute
normalization of the sea and gluon distributions in the pomeron
is determined by one and the same flavor-independent
dimensional constant
$A_{3\Pom}^{*}$ which must approximately be equal to the
triple-pomeron coupling as measured in the real photoproduction.
The normalization of the valence $q\bar{q}$ structure function
of the pomeron is given by the very similar but flavor-dependent
dimensoinal constant (cf. equations (\ref{eq:2.4.1}) and
(\ref{eq:3.3.5},\ref{eq:3.3.7})).
In the above DLLA analysis we omitted the quark loops in the
ladder diagrams for the pomerons. These quark loops shall
automatically be included in the GLDAP-evolution calculation
of the structure function of the pomeron starting with
the above described input parton distributions in the pomeron.



\section{Flux of the QCD pomerons in the proton}

To complete our analysis we must replace the
Low-Nussinov two-gluon pomeron in the lower part of diagrams
of Fig.~7c by the full QCD pomeron - the sum of the triple-ladder
diagrams of Fig.~7d. This is done by replacing the
dipole cross section $\sigma(\rho)$ by $\sigma(y,\rho)$
in the last line of Eq.~(\ref{eq:5.4}),
where $y$ is a fraction of the proton's momentum carried
by the pomeron. Then, the so calculated diffraction dissociation
cross section will differ from that of Section 5 only by the
$y$ dependent factor
\beq
f_{\Pom}(y)=
{\int d\rho^{2}\left[{\sigma(y,\rho)/\rho^{2}}\right]^{2}
{\cal F}(\mu_{G}\rho)   \over
\int d\rho^{2}\left[{\sigma(\rho)/ \rho^{2}}\right]^{2}
{\cal F}(\mu_{G}\rho)  } \, .
\label{eq:6.1}
\endeq
What is the proper interpretatation of $f_{\Pom}(y)$ ?

We would like to preserve the most important result of
the above analysis - the representation of the diffraction
dissociation cross section through the GLDAP-evolving
structure function of the pomeron.
The scaling variable of the photon-pomeron scattering
(\ref{eq:1.3}) equals
$ x_{\Pom}={Q^{2}/(Q^{2}+M^{2})}=x/y$, so that
\beq
{dM^{2} \over M^{2}+Q^{2} } = {dy \over y}
\label{eq:6.2}
\endeq
 With allowance for
the factor $f_{\Pom}(y)$  Eq.~(\ref{eq:6.1})
for the diffraction dissociation cross
section can be written down in the factorized form
\beq
\left.{ d\sigma_{D} \over dt\,dy }\right|_{t=0} =
{  \sigma_{tot}(pp) \over 16\pi}\cdot
{4\pi \alpha_{em}
\over Q^{2}}
{f_{\Pom}(y) \over y} F_{2}^{(\Pom)}({x\over y},Q^{2})\,\, ,
\label{eq:6.3}
\endeq
which has the conventional parton model representation
with $f_{\Pom}(y)/y$ being the flux of pomerons treated as
partons of the proton. In order
not to introduce any spurious dependence on the kinematical
variables $x$ and $y$, the coefficient $\sigma_{tot}(pp)/16\pi$
in (\ref{eq:6.3}) must be taken constant, for instance
fixing $\sigma_{tot}(pp) = 40\,$mb.
Because the pomeron is not the particle with the well defined
couplings (residues) and spin, and because in QCD
the pomeron does not factorize, the
regge-theory convention (\ref{eq:1.1}) is not
unique. The coefficient $\sigma_{tot}(pp)/16\pi$ is the
convention-dependent normalization constant for the
correct dimensionality of the diffraction dissociation
cross section in terms of the dimensionless structure
function or vice versa. The absolute normalizations
of the flux of pomerons and of the pomeron structure
function too are the convention dependent ones: it is always
the product of the two quantities which enters
the experimentally observable cross sections.
Eq.~(\ref{eq:6.3}) shows how the QCD evolution effects and the
rising dipole cross section $\sigma(x,\rho)$ modify the
$1/M^{2}$ law (\ref{eq:1.4}) for the mass spectrum in the
triple-pomeron region.  The factorization (\ref{eq:6.3}) bears
certain semblance of the usual Regge-theory factorization in
the triple-Regge region. We emphasize that we have derived
(\ref{eq:6.3}) neither assuming nor using any of the
Regge-theory factorization relations.

The three pomerons in the triple-pomeron diagram are described
by different QCD ladder diagrams. The top pomeron of
Fig.~7d is in the LLA regime: the relevant sizes vary
along the ladder from $\rho_{n+2}^{2} \sim R_{\Pom}^{2}$ in the
bottom cell of the ladder
down to $r^{2} \sim 1/Q^{2}$ in the top,
quark-antiquark, cell of the ladder. For this reason we
can introduce the structure function of the pomeron.
By contrast, the
exchanged pomerons in the lower part of Fig.~7d
are the soft pomerons: since
$\rho_{n+2}^{2} \sim R_{\Pom}^{2} \sim R_{N}^{2}$,
the sitiuation is reminiscent of
the pomeron contribution to the typical hadronic cross
section, see eq.~(\ref{eq:2.2.3}).
The predictive power of QCD for the hadronic total
cross section is still very limited [44,19,17,45].
(We shall comment more on Lipatov's work on the QCD
pomeron [29] below.) The empirical observation is that
the hadronic cross sections and the real photoabsorption
cross section [46] have a very weak dependence
on energy $\nu \sim m_{N}/x$, much weaker compared to
a steep rise of the DIS structure functions
with $1/x$. The dipole cross section (\ref{eq:4.7}) is
consistent with this property: it is essentially flat
{\sl vs.} $1/x$ at large, hadronic, size $r$, and
the smaller is the size $r$, the steeper is the rise
of $\sigma(x,r)$ with $1/x$.

This rise has a certain impact
on the radius of the pomeron. Namely, the
replacement of $\sigma(r)$ by
$\sigma(x,r)$ leads to the effective wave function of
the pomeron
\beq
|\Psi_{\Pom}(y,z_{G},r)|^{2} =
{81 \over 8  \pi^{4}}
\cdot {1\over z_{G}}\cdot{1\over \sigma_{tot}(pp)}
\left[{\sigma(y,r)\over r^{2}}\right]^{2}
{\cal F}(\mu_{G}r) \, .
\label{eq:6.4}
\endeq
With the rising dipole cross section $\sigma(y,r)$ Eq.~(4.7)
the ratio $\sigma(y,r)/r^{2}$ will rise towards small $r$,
so that the effective radius of the pomeron $R_{\Pom}(y)$
will decrease as $y \rightarrow 0$. The GLDAP evolution
equations hold at
$R_{i}^{2}Q^{2}  \gg 1 $,
where $R_{i}$ are
the relevant hadronic radii. If one would like to formulate
the input for the GLDAP evolution at certain fixed
$Q_{0}^{2}$  then, because of the decreasing radius of the
pomeron, the GLDAP applicability condition will be violated at
very small values of $y \lsim y_{c}(Q_{0}^{2})$ such that
\beq
R_{\Pom}(y_{c}(Q^{2})^{2} Q^{2} = 1 \,.
\label{eq:6.5}
\endeq
The factorization of the diffraction dissociation cross section
into the flux of pomerons and the structure function of pomerons
Eq.~(\ref{eq:6.3}) will break down at $y <y_{c}(Q^{2})$. Our criterion
(\ref{eq:6.5}) for the breaking of the GLDAP evolution is
different from GLR criterion ([19],
for the
review and references see [22])
and its
implications will be studied in the subsequent publications.

     A brief comment on the target dependence is in order. In
the Regge theory, in the approximation of exchange by the
single factorizing  pomeron,
\beq
{1 \over \sigma_{tot}(pp)}
\left.{d\sigma_{D}(\gamma^{*}+p \rightarrow X+p)
\over dtdM^{2}}\right|_{t=0} =
{1 \over \sigma_{tot}(\pi\pi)}
\left.{d\sigma_{D}(\gamma^{*}+\pi \rightarrow X+\pi)
\over dtdM^{2}}\right|_{t=0}\,\,.
\label{eq:6.6}
\endeq
In our $s$-channel approach the target-dependence enters through
the target-dependent dipole cross section (\ref{eq:2.1.4}),
which is roughly proportional to the number of quarks in
in the target hadron $b$ (To the extent that the quark-quark
separation in the proton and the antiquark-quark separation
in the pion are similar, the Low-Nussinov cross section
(\ref{eq:2.1.4}) reproduces the additive quark model
[33,14,15].):
\beq
{\sigma((q\bar{q})b,\rho) \over
 \sigma((q\bar{q})p,\rho)} \approx
{\sigma_{tot}(bp) \over
 \sigma_{tot}(pp)}   \approx
{\sigma_{tot}(b\pi) \over
 \sigma_{tot}(p\pi)}   \, \, ,
\label{eq:6.7}
\endeq
so that despite the lack of factorization, the QCD pomeron
roughly satisfies the relation (\ref{eq:6.6}).



\section
{Unitarization of the rising structre functions}



\subsection
{Rising cross sections and the $s$-channel unitarity}

The rising cross section $\sigma(x,r)$  Eq.~(\ref{eq:4.7})
conflicts the $s$-channel unitarity at sufficiently
large $1/x$. The $s$-channel unitarity constraint is best
formulated in the impact-parameter
representation (the partial-wave expansion) and reads
\beq
\Gamma(b) \leq 1  \, .
\label{eq:7.1.1}
\endeq
The profile function $\Gamma(\vec{b})$ is related to the
elastic scattering amplitude $f(\vec{q})$ such that
\beq
f(\vec{q})=2i\int d^{2}\vec{b}\,
 \Gamma(\vec{b})\exp(-i\vec{q}\,\vec{b})
=i\sigma_{tot}\exp(-{1\over 2}B_{el}\vec{q}\,^{2}) \, .
\label{eq:7.1.2}
\endeq
Here $B_{el}$ is the diffraction slope of the elastic
scattering. The Gaussian parametrization (\ref{eq:7.1.2})
is viable for the purposes of the present discussion [47]
and gives
\beq
\Gamma(b)={\sigma_{tot} \over 4\pi B_{el}}
\exp\left(-{b^{2}\over 2B_{el}}\right) \, .
\label{eq:7.1.3}
\endeq
The profile function of the bare pomeron excahnge $\Gamma_{0}(b)$
defined for the rising cross section (\ref{eq:4.7}) will
overshoot the $s$-channel unitarity bound at a sufficiently
small $x$.

The are no unique prescriptions how to impose the $s$-channel
unitarity constraint on the rising cross sections.
The often used procedures are the eikonal [48,49]
\beq
\Gamma(b)=1-\exp[-\Gamma_{0}(b)] = \sum _{\nu=1}
{(-1)^{\nu-1}\over \nu!}\Gamma_{0}(b)^{\nu}
\label{eq:7.1.4}
\endeq
and the ${\cal K}$-matrix  [50,44]
\beq
\Gamma(b)={1\over 1+\Gamma_{0}(b)} = \sum _{\nu=1}
(-1)^{\nu-1}\Gamma_{0}(b)^{\nu}
\label{eq:7.1.5}
\endeq
$s$-channel unitarizations. Both produce $\Gamma(b)$ which
satisfies the unitarity bound (\ref{eq:7.1.1}).
To the leading order in the $s$-channel unitarization,
the unitarized profile function reads
\beq
\Gamma(b)
\approx \Gamma_{0}(b) - {1 \over 2}\chi\Gamma_{0}(b)^{2}
\label{eq:7.1.6}
\endeq
with $\chi=1$ for the eikonal unitarization, and
$\chi=2$ for the ${\cal K}$-matrix unitarization.
The eikonal unitarization
is being routinely used in
high-energy physics [51] and
sums the $s$-channel iterations of the
bare pomeron exchange (Fig.~9a)
when only the elastic scattering intermediate states are
included in the $s$-channel.
Besides the elastic scattering
states as in Fig.~9a, one must include the
inelastic intermediate states of Fig.~9b, which correspond to the
diffraction dissociation of the target nucleon. These inelastic
intermediate states lead to an enhancement of the double and
higher order rescattering terms in expansions
(\ref{eq:7.1.4},\ref{eq:7.1.5})
[52,49,53].
If one starts with the eikonal unitarization (which is an
assumption), and includes
the corrections for the diffraction dissociation of the
target nucleons, then [53]
\beq
\chi \approx 1 +{\sigma_{D}(p\rightarrow X)\over \sigma_{el}(pp) }
\sim 1.5       \,\, .
\label{eq:7.1.7}
\endeq
In fact, the ${\cal K}$-matrix prescription (\ref{eq:7.1.5}) was
obtained in [44] starting with the eikonal unitarization
of $\pi N$ scattering and
 including the inelastic intermediate states in
the QCD inspired model of the diffraction dissociation of pions.



\subsection
{Shadowing correction to the proton structure function}

Now we shall discuss the unitarization (shadowing, absorption)
effects in DIS,
taking full advantage of the diagonalization of the
$S$-matrix in the $(\vec{\rho},z)$-representation, which allows us
to impose the $s$-channel unitarization
on all the multiparton cross sections
$\sigma{n}(\vec{r},\vec{\rho}_{1},...,\vec{\rho}_{n})$ at all the
values of $\vec{r}$ and $\vec{\rho}_{i}$.
Although the unique $s$-channel unitarization procedure is
lacking, we can still develope a sound phenomenology.
We identify the cross section (\ref{eq:4.7}) which leads
to the GLDAP-evolving structure function
$F_{2}^{(N)}(GLDAP,x,Q^{2})$
with the bare-pomeron exchange. The bare-pomeron structure
function $F_{2}^{(N)}(GLDAP,x,Q^{2})$ is the {\sl linear}
functional of the density of partons in the proton:
\arr
F_{2}^{(p})(GLDAP,x,Q^{2})={Q^{2} \over 4\pi\alpha_{em}}
\sigma_{tot}(GLDAP,x,Q^{2}) \nonumber\\
=
\sum_{i}e_{i}^{2}x\left[q_{i}(GLDAP,x,Q^{2})+
\bar{q}_{i}(GLDAP,x,Q^{2})\right]
\label{eq:7.2.1}
\endarr
The construction of the unitarized photoabsorption cross
section goes as follows. For each Fock state we define
the bare
$\Gamma_{0}(b)$
and the unitarized
$\Gamma(b)$
profile functions
and the bare $\sigma_{0}$ and the
unitarized cross section $\sigma^{(U)}$:
\beq
\sigma^{(U)}= 2\int d^{2}\vec{b}\Gamma(b)=
2\int d^{2}\vec{b}\Gamma_{0}(b) -
 2\int d^{2}\vec{b}[\Gamma_{0}(b)
-\Gamma(b)]=
\sigma_{0}-
\Delta \sigma^{(sh)}
\label{eq:7.2.2}
\endeq

Let us derive the shadowing (unitarity) correction to the
scattering of the $q\bar{q}$ Fock state of the photon.
To the leading order in $\Gamma_{o}(b)$,
Eqs.~(\ref{eq:7.1.6}),(\ref{eq:7.1.3}) give
\beq
\Delta \sigma^{(sh)}(\rho)\approx
\chi{\sigma(\rho)^{2} \over 16\pi B(\rho)} \, ,
\label{eq:7.2.3}
\endeq
and the shadowing correction to the total
photoabsorption cross section equals
\arr
\Delta\sigma_{tot}^{(sh)}(\gamma^{*}N,x,Q^{2})=
\int dz d^{2}\vec{\rho}
|\Psi(z,\rho)|^{2}\Delta\sigma^{(sh)}(\rho)
\approx
\chi\int dz d^{2}\vec{\rho}
|\Psi(z,\rho)|^{2}
{\sigma(\rho)^{2} \over 16\pi B(\rho)} \nonumber\\
=
\chi \int dM^{2}
\int dt{d\sigma_{D}(\gamma^{*} \rightarrow q+\bar{q})
\over  dt dM^{2}}=
\chi \int dM^{2}{1 \over B_{D}(M^{2})}
\left.{d\sigma_{D}(\gamma^{*} \rightarrow q+\bar{q})
\over  dt dM^{2}}\right|_{t=0}\nonumber\\
 =
\chi \int {dM^{2}\over M^{2}+Q^{2}}
 {\sigma_{tot}(pp)\over 16\pi B_{D}(M^{2})}
\sigma_{tot}(\gamma^{*}\Pom\rightarrow q+\bar{q},M^{2}) \,\, .
{}~~~~~~~~~~~~~~
\label{eq:7.2.4}
\endarr
Here $B_{D}(M^{2})$ is the diffraction slope for the diffraction
excitation of the mass $M$. The shadowing correction to the
total photoabsorption cross section equals the diffraction
dissociation cross section times the enhancement parameter
$\chi \approx  (1-2)$. The
generalization of eq.~(\ref{eq:7.2.4}) to interactions of the
higher Fock states of the photon is straightforward.
Making use of eq.~(\ref{eq:6.3}), we
obtain the shadowing correction to the structure function
of the proton
\beq
\Delta F_{2}^{(sh)}(x,Q^{2})=
\chi{\sigma_{tot}(pp) \over 16\pi B_{3\Pom}}
\int_{x}^{y_{m}} {dy\over y}
 f_{\Pom}(y)F_{2}^{(\Pom)}({x\over y}, Q^{2})
{B_{3\Pom} \over B_{D}(M^{2}) }  \, \, .
\label{eq:7.2.5}
\endeq
(The slope $B_{3\Pom}$ and the end-point $y_{m}$ of the
pomeron distribution will be defined below.)
Ignore for a minute
the mass-dependence of the slope $B_{D}(M^{2})$.
Since $F_{2}^{\Pom}(x,Q^{2})$ satisfies the GLDAP evolution, the
convolution representation (\ref{eq:7.2.5}) implies that the
shadowing correction to the proton structure function also
satisfies the GLDAP evolution equations! Experimentally,
in all the hadronic reactions and in the diffraction dissociation
of real photons the slope $B_{D}(M^{2})$ exhibits similar
dependence on the excited mass $M$ [27]:
in the triple-pomeron region the slope
is constant to a good approximation,
\beq
B_{D}(M^{2}) =  B_{3\Pom} \approx {1\over 2}B_{el}(hN)\, ,
\label{eq:7.2.6}
\endeq
whereas in the resonance excitation region
\beq
B_{D}(M^{2})\sim B_{el}(hN)  \, .
\label{eq:7.2.7}
\endeq
In the DIS the counterpart of
excitation of resonances is the excitation of
the $q\bar{q}$ Fock states of the photon, for which
we expect the slope (\ref{eq:7.2.7}), whereas for the
higher Fock states and heavier masses the slope  (\ref{eq:7.2.6})
is more appropriate. These assumptions can be tested at
HERA. Consequently, as compared to the
pomeron structure function as measured
in the diffraction dissociation, in the shadowing structure
function (\ref{eq:7.2.5}) the 'valence' $q\bar{q}$ component
of the pomeron enters
with the suppression factor $B_{3\Pom}/B_{D}(M^{2})\approx 1/2$,
which does not effect the QCD evolution properties.
We conclude, that the unitarized structure
function of the deep inelastic scattering
\beq
F_{2}^{(N)}(x,Q^{2})=
F_{2}^{(N)}(GLDAP,x,Q^{2})-\Delta F_{2}^{(sh)}(x,Q^{2})
\label{eq:7.2.8}
\endeq
satisfies the {\sl linear} GLDAP evolution equation.



\subsection
{Brief phenomenology of the shadowing correction to the
proton structure function}

According to Eq.~(\ref{eq:7.2.4}), the relative
shadowing correction to the proton structure function
equals the fraction $w_{DD}$ of DIS wich goes via diffraction
dissociation of photons times the enhancement parameter $\chi$.
In the diffraction dissociation events the
proton changes its longitudinal momentum $p_{L}$ little,
$\Delta p_{L}/p_{L} = y <y_{m} \lsim 0.1$, and appears in the
final state separated from the hadronic debris of the photon
by the rapidity gap
\beq
\Delta \eta = \log\left({1 \over y}\right)
\label{eq:7.3.1}
\endeq
The standard definition of the diffraction dissociation
corresponds to $\Delta \eta \gsim \Delta\eta_{min}=2-2.5 $.
The maximal
value of the rapidity gap is $\eta_{max}=\log(1/x)$.
The estimate of $w_{DD}$ is particularly easy when the
pomeron and proton structure functions are approximately
constant. In this case $f_{\Pom}(y)=1$, the rapidity gap
distribution is flat which is a signature of the triple-pomeron
mechnaism [1],
and combining
equations (\ref{eq:3.3.11}) and (\ref{eq:7.2.4}) we
find [7]
\beq
w_{DD}(x) \approx
{ A_{3\Pom}(0) \over B_{3\Pom}}\int_{x}^{y_{m}} {dy \over y}=
{ A_{3\Pom}(0) \over B_{3\Pom}}\int_{\Delta \eta_{min}}^{\eta_{max}}
d\Delta \eta =
{ A_{3\Pom}(0) \over B_{3\Pom}}\log\left({y_{m} \over x}\right)\, ,
\label{eq:7.3.2}
\endeq
which is roughly $Q^{2}$-independent.
Numerically, $A_{3\Pom}(0)/B_{3\Pom} \approx 0.03$, and
in Ref.~[7] we gave an  estimate $w_{DD} \sim 0.15$
at $x \sim 10^{-3}$ and $Q^{2} \sim 30$ (GeV/c)$^{2}$.
This prediction is consistent with the first determinations
of $w_{DD}$ by the ZEUS collaboration at HERA [28].

For a somewhat more realistic evaluation of $w_{DD}$
let us assume that
\beq
f_{\Pom}(y) \sim \left(y_{m} \over y\right)^{2\Delta} \, ,
\label{eq:7.3.3}
\endeq
where $\Delta =\alpha_{\Pom}(0)-1 \sim 0.1$, as suggested by
the pomeron phenomenology of the hadronic cross sections [16,17].
We also assume that at small $x$ the struicture functions rise
$\propto (1/x)^{\delta}$ with the same exponent $\delta$ for
the proton and pomeron. The analysis of Ref.~[30] gives
$\delta(Q^{2}) \sim  0.21$ at $Q^{2}=4\,(GeV/c)^{2}$ and
$\delta(Q^{2}) \sim  0.31$ at $Q^{2}=15\,(GeV/c)^{2}$.
The experience with the QCD evolution analysis
suggests that the ratio
$F_{2}^{(\Pom)}(x,Q^{2})/F_{2}^{(N)}(x,Q^{2})$
only will weakly change with $Q^{2}$, so that Eq.~(\ref{eq:3.3.11})
can be used for the relative normalization of the
proton and pomeron structure functions. Then,
\beq
w_{DD}(x) \approx
{ A_{3\Pom}(0) \over B_{3\Pom}}\int_{\Delta \eta_{min}}^{\eta_{max}}
d\Delta \eta \exp[-\gamma(\Delta\eta-\Delta\eta_{min})]=
{A_{3\Pom(0)} \over B_{3\Pom} }\cdot
{1\over \gamma}\left[1-\left({x\over y_{m}}\right)^{\gamma}
\right]  \,\, ,
\label{eq:7.3.4}
\endeq
and  to the extent that
$\gamma =\delta - 2\Delta \ll 1$, and
$\gamma(\eta_{max}-\Delta\eta_{min})\lsim 1$, we have still an
approximately flat rapidity gap distribution and
again obtain
the estimate (\ref{eq:7.3.2})
for $w_{DD}$.
Consequently, we predict rather
large shadowing effect in the proton structure function
\beq
{\Delta F_{2}^{(sh)}(x,Q^{2})  \over
F_{2}^{(p)}(x,Q^{2})} \approx \chi w_{DD}(x)\,\, ,
\label{eq:7.3.5}
\endeq
which persists at all $Q^{2}$. In the kinematical
range of the DIS at HERA the shadowing effect can be as large
as $\sim 30 \%$. The more
detailed phenomenology of the shadowing corrections is
presented in Ref.~[30].



\subsection{Unitarization and
shadowing correction to the parton densities}

Since $F_{2}^{(sh)}(x,Q^{2})$ satisfies the GLDAP evolution,
the shadowing correction to the proton structure
function can be reabsorbed into the modification
of the parton densities in the protons. For instance, the
shadowed density of gluons in the proton shall equal
\beq
g(x,Q_{0}^{2}) =
g(GLDAP,x,Q_{0}^{2})
-\chi {\sigma_{tot}(pp) \over 16\pi B_{3\Pom} }
\int_{x}^{y_{m}}{dy \over y^{2}}f_{\Pom}(y)g_{\Pom}({x\over y})\, .
\label{eq:7.4.1}
\endeq
Similarly, the valence and sea $q\bar{q}$ distributions in
the pomeron will modify the sea quark distribution in the
proton. The detailed phenomenology of the shadowing corrections
to the parton distributions in the proton will be presented
elsewhere [31]. Here we just notice that  whereas the GLDAP
defined parton distributions satisfy the momentum sum rule
\beq
\sum_{p=q,\bar{q},g}\langle x_{p}\rangle =
\int_{0}^{1}dx\,x\left\{g(GLDAP,x,Q^{2})+
\sum_{i}\left[q_{i}(GLDAP,x,Q^{2})+\bar{q}_{i}(GLDAP,x,Q^{2})
\right]\right\} = 1 \,\, ,
\label{eq:7.4.2}
\endeq
because of the shadowing correction
this sum rule does not hold for the experimentally measured
shadowed (unitarized) parton distributions.
A crude estimate of violation of the momentum sum rule
(\ref{eq:7.4.2}) is
\beq
\Delta \Sigma_{x}=1-\sum_{p=q,\bar{q},g}\langle x_{p}\rangle
\sim y_{m}\cdot{A_{3\Pom}^{*} \over B_{3\Pom}} \sim 0.003
\label{eq:7.4.5}
\endeq
With the
$\approx 2\%$ systematic normalization errors in the
most accurate measurements of $F_{2}^{(N)}(x,Q^{2})$,
presently the momentum sum rule can not be tested to better
than $5 \%$ [55].
The concept of
the fusion (recombination) of partons
must be used with much caution. For instance, the
shadowing correction to the density of gluons
Eq.~(\ref{eq:7.4.2})
is not proportional to  $g(x,Q^{2})^{2}$
as it is often stated in the
literature ([19-22], see below Section 7.6).
Indeed, the shadowing term is proportional to $A_{3\Pom}^{*}$,
the integrand of which is $\propto[\sigma(y,r)/r^{2}]^{2}
 \propto  [xg(x,Q_{\Pom}^{2}=1/r^{2})]^{2}$ and the
integration is dominated by large hadronic values of
$r \sim R_{\Pom}$ and small virtualities of the fusing gluons
$Q_{\Pom}^{2} \sim  1/R_{\Pom}^{2}$
[23,24,35].

Because of the shadowing the parton
distributions inferred from the GLDAP evolution analysis of
the DIS structure functions will be different from
the operator-product expansion (OPE)
defined parton distributions, which define the
impulse approximation component $F_{2}^{(N)}(GLDAP,x,Q^{2})$
in Eq.~(\ref{eq:7.2.1}).
To this end, an analogy with the
comparison of the electron-nucleus and proton-nucleus
scattering is instructive:
The elastic $eA$ scattering
is described by the sum of the impulse approximation
diagrams of Fig.~10a and is a {\sl linear} functional of
the nuclear charge density. The $eA$ scattering amplitude
measures the charge of the nucleus which equals the sum of
charges of its constituents (nucleons).
Choosing an  appropriate external field, one can
study the whole sequence of the nuclear matrix elements which
will be sensitive to the momentum distribution of nucleons in
the nucleus. For instance, considering the scattering of the
nucleus on the gravitational center one can derive the momentum
sum rule that the constituent nucleons carry the total momentum
of the nucleus [35]. Under the strong condition that the
scattering in external fields is described by the
impulse approximation, {\sl i.e.,} by the exchnage of the
single quantum of the external field,
having measured the amplitudes of scattering in
a variety of external fields  one can reconstruct the momentum
distribution of nucleons in the nucleus. One would recognize in the
above the standard OPE definition of the parton
densities (for instance, see the textbooks [56]).
In the $pA$ elastic scattering the impulse approximation
amplitude $f_{A}(\vec{q})=Af_{N}(\vec{q})G_{A}(\vec{q})$,
where $G_{A}(\vec{q})$ is the body form factor of the nucleus,
has the profile function $\Gamma_{A}(b) \sim A^{1/3}$ which
grossly overshoots the unitarity bound (\ref{eq:7.1.1}).
Consequently, the $pA$ scattering amplitude is subject to the
large unitarization corrections (Fig.~10b) and is a
{\sl nonlinear} functional of the nuclear matter density.

In this context, the GLDAP approach corresponds to the impulse
approximation and Eq.~(\ref{eq:7.2.1}) gives the {\sl linear}
relationship between the Compton scattering amplitide and the
parton densities. The shadowing term is an apparently
{\sl nonlinear} functional of the density of partons in the
proton, but we have proven that this nonlinearity can be cast
in the form of the renormalization of the parton densities,
with retention of the linear GLDAP evolution properties.



\subsection
{The higher order unitarity corrections and fusion of partons}

The higher order unitarity corrections, {\sl i.e.}, the
multiple-pomeron exchanges Figs.~1b,~1d, 9, do
technically give rise to the photon-miltipomeron interactions,
which cast shadow on the very definition of the photon-pomeron
cross section and the pomerons structure function
Eqs.~(\ref{eq:1.1},\ref{eq:1.2}).
The remarkable observation is that ond can still describe the
diffraction dissociation cross section in terms of the pomeron
structure function and the factorization
representation Eq.~(\ref{eq:6.3}), and
these $\gamma^{*}(n\Pom)$
interactions only
 do slightly modify $f_{\Pom}(y)$ and the
simple relationship (\ref{eq:7.2.5}) between the shadowing
staructure function and the pomeron structure function.
Let us start with the unitarization of the diffraction dissociation
cross section. The $s$-channel iterations of the
QCD pomeron exchange to the left and to the right of the
unitarity cut in  Fig.~11 do separately summ up to the unitarized
dipole cross section. For the $q\bar{q}$ Fock state one must
unitarize $\sigma(y,r)$, for the
$q\bar{q}g_{1}...g_{n}$ Fock states one must unitarize
$\Sigma(y,r)={9\over 4}\sigma(y,r)$.
Barring the $q\bar{q}$ state
the flux of pomerons
$f_{\Pom}^{(D)}(y)$ which enters the diffraction dissociation
cross section must be calculated with the substitution
\beq
\sigma(y,r)^{2} \rightarrow \left({4\over 9}\right)^{2}
\Sigma^{(U)}(y,r)^{2}
\label{eq:7.5.1}
\endeq
in the pomeron wave function (\ref{eq:6.4}), so that
\beq
f_{\Pom}^{(D)}(y)=
\left({4\over 9}\right)^{2} \cdot
{\int dr^{2}\left[{\Sigma^{(U)}(y,r)/r^{2}}\right]^{2}
{\cal F}(\mu_{G}r)   \over
\int dr^{2}\left[{\sigma(r)/r^{2}}\right]^{2}
{\cal F}(\mu_{G}r)  } \, .
\label{eq:7.5.2}
\endeq
Apart from this minor change, the perturbative QCD expansion
(\ref{eq:5.4}) will retain its form.

Similarly, in the case of the shadowing correction, Fig.~12, the
higher order unitarity corrections are accounted for by the
substitution
\beq
\sigma(y,r)^{2}\rightarrow
{ 16\pi B_{3\Pom} \over \chi} \cdot
\left({4\over 9}\right)^{2} \cdot
\Delta\Sigma^{(sh)}(y,r)=
{ 16\pi B_{3\Pom} \over \chi} \cdot
\left({4\over 9}\right)^{2}  \cdot
[\Sigma(y,r)-\Sigma^{(U)}(y,r)]
\label{eq:7.5.3}
\endeq
 so that $\chi f_{\Pom}(y)$ in Eq.~(\ref{eq:7.2.5}) shall be
replaced by
 \beq
\chi f_{\Pom}^{(sh)}(y) =
16\pi B_{3\Pom} \cdot
\left({4\over 9}\right)^{2} \cdot
{\int dr^{2}
\left\{[\Sigma(y,r)-\Sigma^{(U)}(y,r)]/r^{4}\right\}
{\cal F}(\mu_{G}r)   \over
\int dr^{2}\left[{\sigma(r)/ \rho^{2}}\right]^{2}
{\cal F}(\mu_{G}\rho)  } \, .
\label{eq:7.5.4}
\endeq
Again, for the $q\bar{q}$ state one must unitarize $\sigma(y,r)$.
The higher order unitarity corrections make the
two fluxes $f_{\Pom}^{(D)}(y)$ and
$f_{\Pom}^{(sh)}(y)$ slightly
different both in the absolute normalization and in the
$y$-dependence.

Equations (\ref{eq:7.5.2}) and (\ref{eq:7.5.4}) summ in a
very compact form all the multiple-pomeron exchanges in
the $s$-channel (Figs.~11,~12). The origin of this remarkable
result is simple: the interaction cross section of the
$n-$parton Fock state of the photon is dominated by the
spatial extension of the softest gluon, which acts as a
constituent gluon of the pomeron. This corresponds to the
dominance of the multipomeron exchange diagrams of Fig.~13.
Consequently, the
unitarization affects only the normalization of the
pomeron wave function but not the QCD evolution
properties of the pomeron structure function.

The shadowing structure function can be reinterpreted
in terms of the fusion of partons from the overlapping
pomerons emitted by the same nucleon, which reduces the
total density of partons. The fusion of partons from
different nucleons of the nucleus was first introduced
in 1975 [54] and remains a viable mechanism for the nuclear
shadowing in deep inelastic scattering [23,24,35].
However, this interpretation must be taken with the grain of salt.
The bare, GLDAP cross section (\ref{eq:4.11}) is a linear
functional of the density of gluons, the unitarized cross
sections $\sigma^{(U)}(x,r)$, $\Sigma^{(U)}(x,r)$ contain
the terms $\propto (-1)^{n+1}[xg(x,r)]^{n}$ which are
sign-alternating. In more general terms, the multi-gluon
exchange contribution is proportional to the many-gluon
density matrix, the elements of which are not neccessarily
positive defined. Evidently, this quantum-mechanical property
is missed in the probabilistic approach to fusion.


\subsection
{Unitarization and linear GLDAP versus nonlinear GLR evolution
equations}

There was much discussion of the unitarization of rising
structure functions in the framework of
the so-called Gribov-Levin-Ryskin (GLR) nonlinear evolution
equation ([19], for the recent review with many
references see [22]). Here we briefly comment on the origin
of the nonlinear term in the GLR equation, following the
standard derivation of the evolution equations [11-13].
(We only consider $x \ll 1$ of our interest.)
One starts with evaluation of the derivative [12]
\beq
 Q^{2} {\partial \over \partial Q^{2}}F_{2}(x,Q^{2}) \propto
 Q^{2} {\partial \over \partial Q^{2}}
 [Q^{2}\sigma_{tot}(\gamma^{*}N,x,Q^{2})]  \, .
\label{eq:7.6.1}
\endeq
In Section 2 we have decomposed
$Q^{2}\sigma_{tot}(\gamma^{*}N,x,Q^{2})$ into the non-LLA
component (\ref{eq:2.3.2}), which we can neglect, and the
LLA component (\ref{eq:2.3.1}), in which all the explicit
dependence on $Q^{2}$ is concentrated in the integration
limit. This is the crucial point, since taking the derivative
(\ref{eq:7.6.1}) and making use of Eq.~(\ref{eq:4.11}) one
obtains one of the small-$x$ GLDAP equations
\beq
 Q^{2} {\partial \over \partial Q^{2}}
 [x q(x,Q^{2})] \propto
Q^{2}\sigma(1/Q) \propto x\alpha_{S}(Q^{2})g(x,Q^{2})\, ,
\label{eq:7.6.2}
\endeq
in which the both {\sl r.h.s} and {\sl l.h.s} are evaluated
at the same value of $Q^{2}$. Notice, that this
property is a result of
the singular behaviour of the integrand in Eq.~(\ref{eq:2.3.1}).

To the contrary, the integrand of the leading shadowing
correction (\ref{eq:7.2.4},\ref{eq:2.4.1}) is a smooth
function of $\rho$. Furthermore, it is dominated
by the contribution from large $\rho \sim R_{\Pom}$.
For this reason, it would be illegitimate to enforce the
LLA   limit of integration $\rho^{2} > 1/Q^{2}$ in the
shadowing correction (\ref{eq:7.2.4}). If,
nonetheless, one proceeds so, then
differentiation in the first line
of Eq.~(\ref{eq:7.2.4}) will give
\beq
 Q^{2} {\partial \over \partial Q^{2}}
 [Q^{2}\Delta\sigma^{(sh)}(\rho^{2}>
 {1\over Q^{2}},x,Q^{2})] \propto
{1 \over Q^{2}B_{3\Pom}}
[Q^{2}\sigma(1/Q)]^{2} \propto
{1 \over Q^{2}B_{3\Pom}} \alpha_{S}(Q^{2})^{2}
[xg(x,Q^{2})]^{2} \,\, .
\label{eq:7.6.3}
\endeq
The familiar form of the GLR
nonlinear shadowing correction to the
GLDAP equation for the density of gluons [19],
\beq
 Q^{2} {\partial \over \partial Q^{2}}
 [x\Delta g^{(sh)}(x,Q^{2})] \propto
{\alpha_{S}(Q^{2})^{2}  \over Q^{2}B_{3\Pom}}
\int_{x}^{y_{m}}{dy \over y}\,\,
[yg(y,Q^{2})]^{2}             \,\, ,
\label{eq:7.6.4}
\endeq
is different from our result (\ref{eq:7.2.5},\ref{eq:6.3}).
Evidently, neglecting the contribution to the shadowing
term from $\rho^{2} < 1/Q^2$ and/or the
$\propto 1/Q^{2}$ corrections to the leading
form of the wave function in
Eqs.~(\ref{eq:2.4.1},\ref{eq:7.2.4}) can not be justified,
which makes the GLR equation highly questionable.
It is interesting to notice, that Mueller and Qiu [20]
had already have expressed similar doubts on the validity of the
GLR nonlinear equation (see also the recent preprint
by Levin and W\"usthoff [21]). The GLR term is a part of
the $\propto 1/Q^{2}$ corrections to the leading shadowing term
given by our Eq.~(\ref{eq:7.2.5}), see also the above discussion
of the fusion of partons in Section 7.5.




\section{Lipatov's pomeron and diffractive deep
 inelastic scattering}

Lipatov and his collaborators have shown [29,42,57] that
the QCD pomeron of the Regge limit $Q^{2},\,r^{2} = const$,
$1/x \rightarrow \infty$ has the intercept
\beq
\alpha_{\Pom} = 1+\Delta_{\Pom} =
1 +  {12\log{2}\over \pi} \alpha_{S}\sim 1.5\,\, .
\label{eq:8.1}
\endeq
which would have given a very rapidly rising total cross section
\beq
\sigma_{tot} \propto
\left({1\over x}\right)^{\Delta_{\Pom}}  \,\, .
\label{eq:8.2}
\endeq
In our scenario of diffractive DIS we have assumed slow growth
of the dipole cross section (\ref{eq:4.7}) with $1/x$ at large,
hadronic, values of $r \sim R_{N}$. Firstly, this is
consistent with, and supported by, the lack of rapid rise
of $xg(x,Q^{2})$ at small $Q^{2}$ (see Eq.~(\ref{eq:4.11}))
found in the latest QCD analysis of the parton distributions
in the proton [55]. Secondly, this assumption is perfectly
consistent with the Lipatov's theory of the pomeron.

Lipatov's starting point is the lowest order QCD cross section
for the scattering of the two color dipoles
$\vec{r}_{1}$ and $\vec{r}_{2}$
\beq
\sigma_{dd}(\vec{r}_{1},\vec{r}_{2})=
{32 \over 9}\alpha_{S}^{2} \int {d^{2}\vec{k}
\over(k^{2}+\mu_{G}^{2})^{2} }
\left[1-\exp(-i\vec{k}\vec{r}_{1})\right]
\left[1-\exp(i\vec{k}\vec{r}_{2})\right]  \, .
\label{eq:8.3}
\endeq
In the frozen-$\alpha_{S}$ approximation,
the profile function of the dipole-dipole scattering
has the form (here the impact parameter
$\vec{b}$ is defined relative to the centers of the
projectile and target dipoles)
\arr
\Gamma_{dd}(\vec{r}_{1},\vec{r}_{2},\vec{b}) \propto
K_{0}(\mu_{G}|\vec{b}+\vec{s}|)^{2}+
K_{0}(\mu_{G}|\vec{b}-\vec{s}|)^{2}
+
K_{0}(\mu_{G}|\vec{b}+\vec{\Delta}|)^{2}+
K_{0}(\mu_{G}|\vec{b}-\vec{\Delta}|)^{2} \nonumber\\
-K_{0}(\mu_{G}|\vec{b}+\vec{s}|)
K_{0}(\mu_{G}|\vec{b}+\vec{\Delta}|)
-K_{0}(\mu_{G}|\vec{b}+\vec{s}|)
K_{0}(\mu_{G}|\vec{b}-\vec{\Delta}|) ~~~~~~~~~~\nonumber\\
-K_{0}(\mu_{G}|\vec{b}-\vec{s}|)
K_{0}(\mu_{G}|\vec{b}+\vec{\Delta}|)+
-K_{0}(\mu_{G}|\vec{b}-\vec{s}|)
K_{0}(\mu_{G}|\vec{b}-\vec{\Delta}|)\, , ~~~~~~~~~
\label{eq:8.4}
\endarr
where $\vec{s}=(\vec{r}_{1}+\vec{r}_{2})/2$ and
$\vec{\Delta}=(\vec{r}_{1}-\vec{r}_{2})/2$.
In the limit of
\beq
b^{2},r_{1}^{2},r_{2}^{2} \ll {1\over\mu_{G}^{2}}
\label{eq:8.5}
\endeq
one can use $K_{0}(x) \propto \log(x)$. Then, the
dependence on $\mu_{G}$ in Eq.~(\ref{eq:8.4}) disappears
and the dipole-dipole profile function acquires the
conformally invariant form [29]
\beq
\Gamma_{dd}(\vec{r}_{1},\vec{r}_{2},\vec{b})
\propto
\log\left[{
|\vec{b}+\vec{\Delta}|
|\vec{b}-\vec{\Delta}|
\over
|\vec{b}+\vec{s}|
|\vec{b}-\vec{s}|}\right]
\log\left[{
|\vec{b}+\vec{\Delta}|
|\vec{b}-\vec{\Delta}|
\over
|\vec{r}_{1}|
|\vec{r}_{2}|}\right] \,\, .
\label{eq:8.6}
\endeq
It differs from (\ref{eq:8.4}) by terms which give vanishing
contribution to the dipole-dipole cross section after the
$d^{2}\vec{b}$ integration. In the same conformal-invariant
limit, also the wave functions of the many body Fock states
will become the scale invariant functions,
see Eq.~ (\ref{eq:3.2.9}).
Because of
this scale invariance no specific size ordering dominates
in the higher order QCD diagrams. Using
the powerful technique of the conformal
field theories, Lipatov has found that with allowance for the
running QCD coupling the QCD pomeron corresponds to the
series of poles, which accumulate at $j=1$ in the complex
angular momentum plane, with the
intercept of the rightmost singularity given by
eq.~(\ref{eq:8.1}). Calculation of residues of these Lipatov's
poles is as yet lacking.

In the scenario of Ref.~[44] the dominant, constant, component
of the hadronic cross section comes from the $j=1$ cross section
(\ref{eq:2.1.4}), whereas the contribution of poles with
$\Delta_{\Pom} > 0$ has a small residue.
Indeed, the dipole cross section $\sigma(\rho)$
is given by the expectation value of the dipole-dipole
cross section over the target nucleon state,
\beq
\sigma(r)={3\over 2}\int d^{3}\vec{R}_{1} d^{3}\vec{R}_{2}
|\Psi_{N}(\vec{R}_{1},\vec{R}_{2},\vec{R}_{3})|^{2}
\sigma_{dd}(\vec{r},\vec{R}_{1\perp}-\vec{R}_{2\perp}) \,\, ,
\label{eq:8.7}
\endeq
and receives the dominant contribution from the separation of
quarks in the proton in
the nonscaling region of
$(\vec{R}_{1\perp}-\vec{R}_{2\perp})^{2}
\sim R_{N}^{2},1/\mu_{G}^{2}$. If Lipatov's conformal pomeron
with the intercept (\ref{eq:8.1}) only is applicable in the
scaling region of
$(\vec{R}_{1\perp}-\vec{R}_{2\perp})^{2}
\ll  R_{N}^{2},1/\mu_{G}^{2}$, then Eq.~(\ref{eq:8.7}) gives a
natural explanation for the small residue of the rightmost
pole, corroborating the scenario [44] (for the evaluation
of the effective intercept of the pomeron in the hadronic
scattering at finite $\mu_{G}$ see Ref. [43]).

Lipatov's analysis and the intercept (\ref{eq:8.1}) with the
running coupling $\alpha_{S}(t)$ are expected to be directly
applicable to the hard elastic $pp,\,\bar{p}p$ scattering
at $|t |\gg m^{2}, \,|t|\ll s$. The dominance of the rightmost
pole of the pomeron would have implied
\beq
{d\sigma_{el}(pp) \over dt} \propto s^{2(\alpha_{\Pom}(t)-1)}
\sim s^{1} \,\, .
\label{eq:8.8}
\endeq
The experimental lack of such a rapid growth of
the differential cross section of elastic scattering with energy
is another strong evidence for the small residue of poles
with $\Delta_{\Pom} > 0$.

The DLLA pomeron and Lipatov's pomeron summ (in a somewhat
different kinematical regime) the
gluon ladders with two gluons in the $t$-channel. They both
share the unpleasant property of running into conflict with
the $s$-channel unitarity, which shows that the initial DLLA
and/or LLA selection of the subset of the perturbative QCD
diagrams for the pomeron is unsatisfactory and open to
a criticism. As a matter of fact, such a construction of the
pomeron lacks self-consistency, since the
sub-leading unitarization corrections do become larger
than the 'leading' single-pomeron exchange. From the numerical
point of view, the above presented phenomenology seems to be
viable in the kinematical range of HERA:
according to our analysis in Section 7.3, the
unitarization effects are
numerically large, but still significantly smaller than
the leading GLDAP component of the structure function (for the
more detailed phenomenology of the shadowing see Ref.~[30]).
More theoretical work on unitarization is badly needed.
Some 16 years ago, Matinyan and
Sedrakyan [58] have ponted out the potential significance of the
mutiparticle Regge singularities (Fig.14). In the recent very
interesting paper [59], Bartels has shown that such a correlated
four-gluon exchange leads to the new singularities in the
complex angular momentum plane which are missed when the
eikonal-like exchange by the two rightmost
 poles of the Lipatov's pomeron is
considered. The phenomenological implications of this result
for the shadowing corrections to the proton structure functions
are not yet clear, since the quantitative
understanding of residues of Lipatov's poles
in the DIS regime is lacking. Our minimal-regularization
approach, in which the radiative generation of the glue in
protons ([35,37], see also [39])
is closely related to the slow rise of $\sigma(x,r)$ at large,
hadronic, values of $r$,
leads to a good parameter-free description of the proton
structure function statring with the SLAC-NMC range of $x$
and $Q^{2}$ [23,24,35] and down to the kinematical range of
HERA, where we have obtained [30] a good agreement with
the HERA results on $F_{2}^{(p)}(x,Q^{2})$ [60].


\section{Conclusions and discussion of results}

Our principle conclusion is that the diffraction dissociation
of virtual photons in DIS can be described as the DIS on
pomerons with the well-defined and GLDAP-evolving structure
function.
Our analysis completes
the proof of the parton model phenomenology of the pomeron put
forward some 8 years ago by Ingelman and Schlein [2].
Furthermore, we have shown that
such a description persists beyond the single-pomeron
exchange approximation. Our new result is that we
have identified the valence $q\bar{q}$,
the valence glue and the sea $q\bar{q}$ parton distributions
in the pomeron, which are to be used as an input in the QCD
evolution of the pomeron structure function.
We have found that the normalization
of the valence glue and sea in the pomeron is fixed by the
single dimensional coupling $A_{3\Pom}^{*}$, which
is sensitive to the infrared regularization.
Our principle finding is that this coupling $A_{3\Pom}^{*}$
(and the corresponding triple-pomeron
 coupling $A_{3\Pom}(Q^{2})$ which we have shown
only weakly depends on $Q^{2}$) must be approximately equal
to the triple-pomeron coupling $A_{3\Pom}(0)$ as measured in
the diffraction dissociation of the real photons [27].
This approximate equality $A_{3\Pom}^{*} \approx A_{3\Pom}(0)$
was conjectured long ago [26] and has been a basis of the
successful phenomenology of the nuclear shadowing in DIS
[23,24,35]. This equality also was used in the prediction [7]
of the rate of the diffraction dissociation in DIS, which
is in good agreement with the first data by the ZEUS
collaboration [46].
Important implication of separation of
the infrared-sensitive input structure
function of the pomeron from the hard QCD evolution effects
is that the jet activity in the DIS on the pomeron must be
similar to that in the DIS on the proton.

We have derived the unitarity
(shadowing) correction to the proton structure function
at small $x$ and have demonstrated, that the unitarized
structure function satisfies the conventional, linear, GLDAP
evolution equations. We emphasize the intrinsic simplicity
of our light-cone $s$-channel formalism  used in this derivation.
Firstly, our
formalism implements in a very simple way the color gauge
invariance constraints. Secondly, {\sl exact} factorization of the
photoabsorption cross section into the wave function and
into the (multiparticle) dipole cross section allows an easy
identification of the partial waves of
the dipole cross section as an  object of
the $s$-channel unitarization. Thirdly, we took full advantage
of the diagonalization of the scattering matrix as a function
of the transverse separation and longitudinal momenta of
partons in the multiparton Fock states of the photon. This
enabled us to easily impose the $s$-channel unitarization on
the total cross sections of all the multiparticle Fock states
of the photon. This also enabled us to identify the
constituent gluon wave function of the pomeron, which
gives a very economic description of the shadowing process in
terms of the single parameter $A_{3\Pom}^{*}$, which is under
the good control as it is related to the triple-pomeron coupling
$A_{3\Pom}(0)$ known from the real photoproduction experiments.
We have shown how the multipomeron exchanges in the shadowing
structure function and in the diffraction dissociation can be
summed in a very compact form which only renormalizes the
effective flux of pomerons in the proton.

Applications of the above formalism to the hadronic scattering
problem will be presented elsewhere [43].\bigskip\\

\centerline{\bf \Large Acknowledgements}
One of the authors (NNN)
would like to thank  V.V.Anisovich and L.G.Dakhno for
criticism and L.N.Lipatov for numerous discussions on the
pomeron and deep inelastic scattering during the past years.
We are grateful to V.Barone, M.Genovese, E.Predazzi and
V.R.Zoller for useful comments and discussions. \bigskip\\

\pagebreak

\pagebreak
{\bf \Large Figure captions}
\begin{itemize}
\item[Fig.1]
 - The single- and multiple-pomeron exchange contributions
to the {\sl (a,b)} elastic scattering and {\sl (c,d)} the
diffraction dissociation amplitudes.

\item[Fig.2]
 - The lowest order QCD diagrams for interaction
  of the $q\bar{q}$ Fock state of the photon with the target
  nucleon. In all the figures the wavy, solid and dashed lines
are for the photon, (anti)quarks and gluons, respectively.

\item[Fig.3]
- One of the 16 lowest order QCD diagrams for the inclusive cross
  section of the forward diffraction dissiociation of the
  $q\bar{q}$ Fock state of the photon.  The vertical dashed
  lines shows the unitarity cut corresponding to the
  diffractively excited state.

\item[Fig.4]
- The spatial structure of the $q\bar{q}g$ Fock state in the
  impact-parameter plane.

\item[Fig.5]
- Scattering of the $q\bar{q}g$ Fock state of the photon on
  the nucleon by interaction of its radiatively generated gluon.

\item[Fig.6]
- Different couplings of the exchanged gluons to the
  color-octer $q\bar{q}$ pair and the gluon of the $q\bar{q}g$
  Fock state of the photon.

\item[Fig.7]
   The triple-pomeron diagrams for the diffraction dissociation
of virtual photons in the deep inelastic scattering:
 \subitem[a]
 - The triple-pomeron diagram which describes the
   $\propto 1/M^{2}$ component of the mass spectrum in the
   triple-Regge phenomenology of diffraction dissociation.

 \subitem[b]
 - The driving term of the triple-pomeron mass spectrum in QCD -
   the diffraction excitation of the
   $q\bar{q}g$ Fock state of the photon.

 \subitem[c]
 - Diffraction excitation of the many-particle Fock states
   in the Low-Nussinov approximation for the exchanged pomerons.

 \subitem[d]
 - The same as $(c)$ with the exchange by the full QCD pomerons.

\item[Fig.8]
 - Gluon-ladder representation of the DLLA pomeron in QCD.

\item[Fig.9]
 - $s$ channel iteration of the pomeron exchange:
\subitem[a]
 - In the approximation of elastic intermediate states.
\subitem[b]
 - Contribution of the inelastic intermediate states (diffraction
   dissociation of the target) to the $s$-channel iteration of
   the pomeron exchange.

\item[Fig.10]
 - The archetype operator product expansion:
\subitem[a]
 - The
 impulse approximation diagram for the electron-nucleus scattering
 which is a linear probe of the nuclear charge distribution.
\subitem[b]
 - Multiple scattering diagrams which unitarize the proton-nucleus
 elastic scattering amplitude.

\item[Fig.11]
 - Absorption (unitarization)
 corrections to the diffraction excitation of the
 $q\bar{q}g_{1}....g_{n}$ Fock state of the photon.
    The vertical dashed line shows the unitarity cut.

\item[Fig.12]
 - Unitairzation of the scattering amplitude for the
 $q\bar{q}g_{1}...g_{n}$ Fock state of the photon by the $s$-channel
 iteration of the QCD pomeron exchange.

\item[Fig.13]
 - The dominant multipomeron interactions in the deep
inelastic scattering.

\item[Fig.14]
 - The multiparticle Reggeons.

\end{itemize}
\end{document}